\documentclass[journal]{IEEEtran}
\usepackage[T1]{fontenc}

\ifCLASSINFOpdf
\usepackage[pdftex]{graphicx}
\DeclareGraphicsExtensions{.pdf,.jpeg,.png}
\else
\fi

\usepackage{cite}
\usepackage{graphicx}
\usepackage[cmex10]{amsmath}
\usepackage{algorithm}
\usepackage{algorithmic}
\usepackage{bm}

\usepackage{amssymb}
\usepackage{breqn}
\usepackage{array}
\usepackage{url}
\usepackage{placeins}
\usepackage{multirow}
\usepackage{lipsum}
\usepackage{mathtools, cuted}
\usepackage{multicol, blindtext}

\usepackage{subfloat}
\usepackage{caption}
\usepackage{subcaption}
\usepackage{amsthm}

\renewcommand{\baselinestretch}{1.0}

\def\BibTeX{{\rm B\kern-.05em{\sc i\kern-.025em b}\kern-.08em
    T\kern-.1667em\lower.7ex\hbox{E}\kern-.125emX}}
\begin{document}
\title{Low-Resolution ADC Quantized Full-Duplex Massive MIMO-Enabled Wireless Backhaul in Heterogeneous Networks Over Rician Channels }

\author{\IEEEauthorblockN{Prince~Anokye,~\IEEEmembership{Member,~IEEE,}
        Roger Kwao~Ahiadormey,~\IEEEmembership{Student Member,~IEEE,}
        Han-Shin~Jo,~\IEEEmembership{Member,~IEEE,}
        Changick~Song,~\IEEEmembership{Senior Member,~IEEE,}
        and~Kyoung-Jae~Lee,~\IEEEmembership{Member,~IEEE}}% <-this % stops a space
\thanks{This work was supported in part by the National Research Foundation of Korea (NRF) grant funded by the Korea government (MSIT) (2019R1A2C4070361), and in part by the Institute for Information and Communications Technology Promotion (IITP) Grant funded by the Korea Government (MIST) (2016-0-00500, Cross layer design of cryptography and physical layer security for IoT networks).

P. Anokye, R. K. Ahiadormey, H.-S. Jo, and K.-J. Lee are with the Department of Electronic and Control Engineering, Hanbat National University, Daejeon 34158, Republic of Korea (email: princemcanokye@yahoo.com, rogerkwao@gmail.com, hsjo@hanbat.ac.kr, kyoungjae@hanbat.ac.kr).
C. Song is with the Department of Electronic Engineering, Korea
National University of Transportation, Chungju 27469, Republic of Korea (e-mail:
c.song@ut.ac.kr).
Corresponding author: Kyoung-Jae Lee (kyoungjae@hanbat.ac.kr).
}
}

% make the title area
\maketitle

\begin{abstract}
This paper studies the spectral/energy efficiency (SE/EE) of a  heterogeneous network with the backhaul enabled by low-resolution analog-to-digital converters (ADCs) quantized full-duplex massive multiple-input multiple-output (MIMO) over Rician channels. Backhaul communication is completed over two phases. During the first phase, the macro-cell (MC) base station (BS) deploys massive receive antennas and a few transmit antennas; the small-cell (SC) BSs employ large-scale receive antennas and a single transmit antenna. For the second phase, the roles of the transmit and receive antennas are switched. Due to the low-resolution ADCs, we account for quantization noise (QN).  
We characterize the joint impact of the number of antennas, self-interference, SC-to-SC interference, QN, and Rician $K$-factor.  
For the first phase,  the  SE is enhanced with the massive receive antennas and the loss due to QN is limited. For the second phase,  the desired signal and QN have the same order. Therefore, the SE saturates with the massive transmit antennas. 
As the Rician $K$-factor increases,  the SE converges. Power scaling laws are derived to demonstrate that the transmit power can be scaled down proportionally to the massive antennas. 
We investigate the EE/SE trade-offs. 
The envelope of the EE/SE region grows with increase in the Rician $K$-factor.      
\end{abstract}

% Note that keywords are not normally used for peerreview papers.
\IEEEpeerreviewmaketitle

\section{Introduction}  
\IEEEPARstart{T}{o} make efficient use of the  wireless spectrum with high energy efficiency (EE), full-duplex (FD), massive multiple-input multiple-output (MIMO),  and heterogeneous  networks (HetNets) have been identified as enabling technologies for the next-generation communication systems. In-band FD  promises to double the spectral efficiency (SE) relative to half-duplex (HD) communication by permitting signal transmission and reception on the same frequency band \cite{Duarte10}. However, this SE enhancement is constrained by the self-interference (SI), i.e., the base station (BS) receives the signal leakage  from its own transmissions. Using omni-directional and directional antennas, \cite{Duarte10} and \cite{Duarte14} show that SI can be significantly suppressed. The authors of \cite{Bliss07,Riihonen11,Riihonen09} proposed multi-antenna techniques such as zero-forcing (ZF) beamforming, minimum mean square error (MMSE) filters, and null-space projection to cancel the SI. 
%Despite the deleterious impacts of SI, recent works have shown that FD is feasible \cite{Kim:15}.
%
Of immense interest also is massive MIMO, where the BSs employ far higher number of antennas  than the user terminals (UTs) \cite{Marzetta10}. The large degree of freedom  enables the excellent suppression of noncoherent interference, fast fading, and noise   while improving the SE and EE \cite{Marzetta10}. 
%Marzetta showed in \cite{Marzetta:10} that in the large-scale antenna regime, simple linear processing techniques such as maximum ratio combining/transmission (MRC/MRT) achieve high SE. 
The pilot contamination  which introduces a performance ceiling in the multi-cell massive MIMO has been solved by using the multi-cell MMSE precoding/combining in \cite{Bjornson18}.  Harnessing the benefits of both massive MIMO and FD, several papers have combined the two technologies \cite{Ngo14, Wang17, Sho17}.  The  authors of \cite{Ngo14} and \cite{Wang17} studied a multi-pair decode-and-forward (DF) relay employing  FD massive MIMO and showed that as the number of antennas tends to infinity, the SI is asymptotically canceled.   In \cite{Sho17}, an SI-aware downlink (DL) precoder and uplink (UL) receive filter are proposed. 

%A hardware impairment aware transceiver is proposed for massive MIMO FD relays in \cite{Xia:15}.

HetNets which involve densifying a high-powered macro-cell (MC) BS with a number of low-powered small-cell (SC) access points have been shown to provide higher SE and EE relative to traditional cellular networks  \cite{Sang15}. In HetNets, the MC BS provides coverage for medium to high-mobility UTs whereas the SC BSs serve low-mobility to stationary UTs. In contrast to conventional systems, where the BSs have stable and reliable wired backhaul connections, in HetNets, due to the varying requirements of error rates, delay, capacity, and installation costs associated with each SC BS, the backhaul becomes the main limitation \cite{Sang15}. Thus, an economically viable approach would be to leverage the available radio resources for the backhaul connections. The UL/DL power consumption of a wireless backhaul is investigated in \cite{Sang15}, where the MC BS employs massive antennas to serve mobile users and provides backhaul support for the single-antenna SC BSs serving stationary users. Wang \textit{et al.} \cite{Wang16} studied a sum logarithmic user rate maximization problem for a massive MIMO-enabled backhaul. Authors of \cite{Xia16} optimized the bandwidth allocation between the access and backhaul links, where the MC BS employs massive MIMO and the SC BSs utilize single antennas. Anokye \textit{et al.} in \cite{Anokye17}  proposed a backhaul topology for a two-tier HetNet, where FD  massive MIMO is employed not only at the MC BS but also at the SC BSs. Extending the results in \cite{Anokye17}, \cite{Anokye18} optimizes the pilot length with the aim of maximizing the sum-rate and proposes a hybrid FD/HD architecture. 
%Here, a closed-form solution to operate the hybrid system is derived. 
%The pilot length is optimized with the aim of maximizing the sum-rate. 
%Using tools from stochastic geometry, Tabassum \cite{Tabassum:16} studied massive MIMO-enabled backhaul, where the single-antenna SC BSs operate either in FD or HD mode. 
%Here,  the optimal  ratio of the FD SCs to the HD SCs is derived.

Inevitably, the use of large-scale antennas leads to a significant growth in the power consumption and hardware cost since each antenna requires a pair of high-resolution analog-to-digital converters (ADCs) (e.g. 10-12 bits for commercial applications) \cite{Zhang2017}. A $b$-bit ADC with a sampling frequency $f$ makes $2^b\cdot f$ computations per second. Thus the power consumption grows exponentially with the resolution  and linearly with the sampling frequency. 
To alleviate this challenge, low-resolution ADCs (e.g., 1-3 bits) have been proposed for massive MIMO systems.   However, using low-resolution ADCs introduces quantization noise (QN) into the received signal. Mo and Heath \cite{Mo18}, under the assumption of perfect channel state information (CSI) and finite-bit ADCs, proposed a codebook design for multi-user multiple-input single-out (MISO) channels in limited feedback systems.  
An UL  throughput analysis of massive MIMO with low-resolution ADCs is studied in \cite{Jacobsson17}. In \cite{Dong17}, the authors studied a multi-user massive MIMO  amplify-and-forward (AF) relay with low-resolution ADCs and \cite{Kong17} investigates an FD massive MIMO AF relaying system. The authors of \cite{Liang16} investigated a mixed-ADC massive MIMO, where a few number of high-resolution ADCs are reserved for channel estimation with 1-bit ADCs utilized for data reception. Here, the authors developed an optimal ADC switch and studied the  performance under the generalized mutual information framework.  Considering a mixed ADC/digital-to-analog (DAC) multipair AF massive MIMO relaying, \cite{Zhang2019} analyzes the sum-rate and proposes a power allocation scheme with the aim of maximizing the sum-rate.   An UL SE of an HD massive MIMO  with low-resolution ADCs and an SE/EE trade-off analysis in a mixed ADC scenario are investigated in \cite{JZhang16} and \cite{Zhang2017}, respectively,  over Rician channels.  

The analyses of massive MIMO and/or FD-enabled backhauls in HetNets have predominantly ignored the role of the ADCs. 
Considering that a higher number of SC BSs (and by extension, more receive antennas) are used in HetNets compared to traditional systems, the hardware cost and power consumption of an FD massive MIMO-enabled HetNet with high-resolution ADCs would be prohibitive. Therefore, for a practical deployment, it is crucial to study the feasibility of  low-resolution ADCs. Although, there are a number of researches on low-resolution ADCs in other massive MIMO scenarios \cite{Mo18, Dong17, Jacobsson17, Kong17, Zhang2017,Zhang2019}, the results are not directly applicable to the massive MIMO-enabled backhaul. The authors of \cite{Kyoungjae18} studied  an FD massive MIMO-enabled backhaul with low-resolution ADCs employed at the MC BS and SC BSs. However, it was limited to a Rayleigh fading and ignored the impact of the ADC power consumption on the EE.  
%Furthermore, the work in \cite{Kyoungjae:18} raises concerns of practical implementation due to the increasing physical size of the massive arrays in the SC BSs. Millimeter wave offers a way out, since the  wavelength allows the massive antenna elements to be packed in a relatively small physical volume. 
%For millimeter waves, the assumption of Rayleigh fading fails to  capture the dominant line-of-sight (LoS) components \cite{Zhang:14}.

\begin{figure*}[t!]
\centering
\includegraphics[width=14.0cm,height=12.0cm,keepaspectratio]{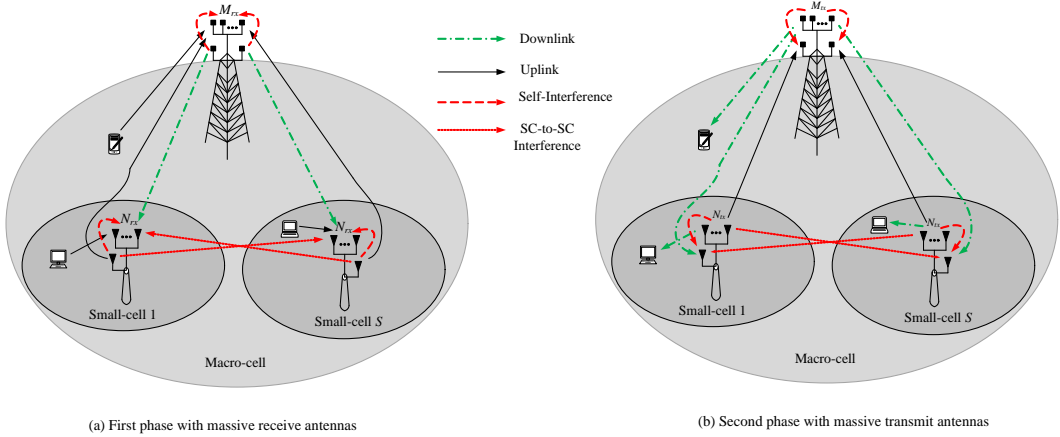}
\caption{HetNet system model with FD massive MIMO wireless backhauls.}
\label{figSysModel}%\vspace{-0.4cm}
\end{figure*}

Specifically, this paper studies a HetNet backhaul  link which is supported by low-resolution ADC quantized FD massive MIMO over Rician fading channels. Data communication is completed in two phases. During the first phase, the MC BS is equipped with massive receive antennas to receive information from the SC BSs and a few transmit antennas (i.e., fixed to be equal to the number of SC BSs)  to send data to a corresponding SC BS in the DL. 
On the other hand, each SC BS employs massive receive antennas and a single transmit antenna to send data to the MC BS in the UL (see Fig. \ref{figSysModel}(a)). 
\begin{figure}[!htbp]%\vspace{-0.2cm}
\centering
\includegraphics[width=8.5cm]{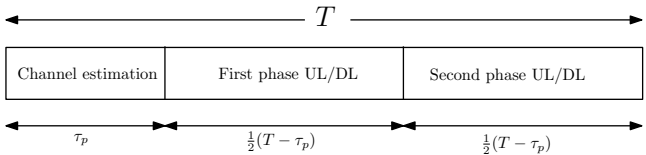}%\vspace{-0.4cm}
\caption{FD massive MIMO frame structure with coherence interval $T$.}
\label{figcoher}%\vspace{-0.4cm}
\end{figure}
For the second phase, the roles of the transmit and receive antennas are switched with the aid of a circulator \cite{Carchon2000} such that the MC BS now has massive transmit antennas and a few receive antennas (equal to the SCs). Here, the SC BSs now  possess massive transmit antennas and single receive antennas to receive independent stream from the MC BS (see Fig. \ref{figSysModel}(b)). All the BSs are FD-capable and therefore transmit and receive on the same frequency band during both phases.\footnote{This configuration enables the BSs to serve UL users during the first phase with the massive receive antennas and support DL users with the massive transmit antennas in the second phase. Therefore, by design, the access users operate in the HD mode while the MC BS and SC BSs operate in FD. However, we concentrate exclusively on the backhaul link design of the HetNet to highlight the advantages of employing the FD operation with massive transmit or receive antennas compared to the conventional HD based configurations. While it is obvious that the inclusion of access users would introduce more interferences, it is expected that with the massive receive antennas  (in the first phase) or  massive transmit antennas (in the second phase), we can suppress these interferences. Future work will consider the influence of the access users.} Different from \cite{Anokye17,Anokye18} (implicitly assume high-resolution ADCs) and \cite{Kyoungjae18} (which considers low-resolution ADCs), where the  HetNet backhaul is analyzed under the assumption of Rayleigh distributed channels, this paper considers Rician fading which is more general. Again, for millimeter wave propagation, in which massive MIMO can find a plethora of applications, the assumption of Rayleigh fading fails to capture the dominant line-of-sight (LoS) components present in millimeter waves \cite{Zhang14}.   
We characterize the joint impact of the Rician $K$-factor, SI, SC-to-SC interference (occurs due to the FD operation at all the SC BSs), and QN on the SE/EE.   
The main contributions of this paper are summarized as follows:
\begin{itemize}
\item Closed-form solutions are derived for the UL/DL SE of the backhaul under the assumption of imperfect CSI and low-resolution ADCs over  Rician  channels.  For the first phase, the sum SE is generally enhanced with the number of receive antennas. We show that the SE loss due to QN is very limited. 
In contrast, in the second phase,  the SE saturates rapidly with the massive transmit antennas since the desired signal and the QN have the same order. 
%Therefore, we recommend using high-resolution ADCs. This is feasible since far fewer number of receive antennas are utilized in the second phase.
\item We show that as the Rician $K$-factor increases, the SE improves until it converges to a fixed value.  A novel expression for the saturation point is derived.

\item Using the proposed model, we obtain intuitive power scaling laws that guarantee a non-vanishing sum SE as the number of massive receive antennas (in the first phase) tend to infinity, where it is revealed that the low-resolution ADCs at the BSs' receivers do not cause substantial impact on the power scaling, relative to the infinite-resolution ADCs' case. However, for the second phase, due to the fact that the desired signal and the QN have equal order, the sum SE of the low-resolution ADCs' case saturates more rapidly compared to the infinite-resolution ADCs counterpart when we apply the power scaling laws. 

\item Although, the increase in the ADCs' resolution improves the SE in a logarithmic scale, the power consumption of the ADCs grows exponentially with the resolution. We study the SE/EE trade-off as a function of the quantization bits, Rician $K$-factor, massive receive antennas (first phase), and massive transmit antennas (second phase). 
\end{itemize}

The rest of the paper is organized as follows: Section \ref{secsystem} presents the system model and the SE analysis is discussed in Section \ref{secrate}. In Section \ref{seclargesysanalysis}, we derive closed-form solutions for the UL/DL SE and present a SE performance evaluation. In Section \ref{secresults},  we investigate the EE performance under practical power consumption model. Furthermore, we provide numerical results to illustrate the sum SE and EE performances of the low-resolution ADC quantized FD massive MIMO-enabled backhaul topology. Section \ref{secconc} concludes the paper. 
\textit{Notations}: Boldface lower and upper case letters denote vectors and matrices, respectively. $(\cdot)^H$, $(\cdot)^T$, $\mathbb{E}[\cdot]$, and $\text{tr}(\cdot)$ indicate the conjugate transpose, transpose, expectation, and trace operators, respectively. $\text{diag}(\mathbf{A})$ returns the diagonal elements of $\mathbf{A}$ and $[\mathbf{A}]_{mn}$ is the $(m,n)$-th entry of $\mathbf{A}$. $\mathbf{x}\sim \mathbb{CN}(\mathbf{0},\mathbf{N})$ represents a circularly symmetric complex Gaussian vector $\mathbf{x}$ with zero mean and covariance $\mathbf{N}$.

\section{System Model}\label{secsystem}
Consider the two-tier HetNet shown in Fig. \ref{figSysModel}, where a high-powered MC BS is overlaid with a number of low-powered SC BSs $S$. The MC BS has a dedicated wired backbone but connects to the SC BSs through a wireless backhaul link. 
Data communication is completed in two phases. 
During the first phase, the MC BS is equipped with massive receive antennas $M_{rx}$ and a few transmit antennas $M_{tx}$ such that $M_{tx}\ll M_{rx}$. Here, the number of transmit antenas are fixed such that each antenna is dedicated to send independent data to a corresponding SC BS. The SC BSs also deploy massive receive antennas $N_{rx}$ and a single transmit antenna. 
For the second phase, we switch the roles of the antennas using a circulator. 
Thus, the MC BS now possesses massive transmit antennas $M_{tx}$ and a
few receive antennas $M_{rx}$; each SC BS has
massive transmit antennas $N_{tx}$ and a single receive antenna.\footnote{This can be treated as a special case of the MC BS using multiple antennas to transmit multiple streams of data to each SC BS during the first phase and receiving with multiple antennas for the second phase. The same argument holds for each SC BS. It is shown in \cite{Anokye18} that in the Rayleigh fading, the SI strength at the MC BS increases with the number of transmit antennas. This conclusion also holds in the Rician case as will be shown in this paper. Therefore, it is expected that using multiple transmit antennas at the SC BSs could exacerbate the SI and SC-to-SC interference. Future work would investigate this problem.} All the receive antennas have low-resolution ADCs  to quantize the received signal during the data transmission phase. We assume the frame structure shown in Fig. \ref{figcoher}, where $T$ denotes the coherence interval (in symbols) and $\tau_p$ indicates the timeslots used for pilot signaling. The remaining timeslots, i.e., $T-\tau_p$, are then utilized for the first and second phases UL/DL. 
A time division duplex (TDD) protocol is assumed such that the channel estimates used for the first phase signal detection are also employed to precode data in the second phase, i.e., we assume the hardware calibration is perfect such that reciprocity holds \cite{Larsson14}. Furthermore, we use the general assumption that the transmit signals $\mathbf{x}$ and $\mathbf{s}$  follow Gaussian distribution \cite{Zhang2019}.
\subsection{First Phase Data Communication}
The MC BS sends independent signal to the $S$ SC BSs in the DL.
Simultaneously, the $k$-th SC BS sends its signal to the MC BS in the UL. Due to the simultaneous transmission and reception, the MC BS and the SC BSs suffer from SI. In addition to the SI, the SC BSs are affected by  the SC-to-SC interference which occurs due to the FD operation employed at the SC BSs. The received signal at the MC BS and the $k$-th SC BS are expressed, respectively, as
\begin{align}\vspace{-0.3cm}
\mathbf{y}_m^{(1)} = \sqrt{p_s}\mathbf{Hx} + \sqrt{p_m}\mathbf{Qs} + \mathbf{n}\hspace{0.1cm},
\end{align} 
%\vspace{-0.9cm}
\begin{align}
\mathbf{y}^{(1)}_{s,k} &= \sqrt{p_m}\mathbf{g}_{k}s_{k} + \sqrt{p_m}\sum^S_{j=1, j\neq k}\mathbf{g}_{j}s_{j} + \sqrt{p_s}\mathbf{q}_{s,k}x_{k}\nonumber \\&+ \sqrt{p_s}\sum^S_{j=1, j\neq k}\mathbf{q}_{c,kj}x_{j} + \mathbf{v}_{k}\hspace{0.1cm},
\end{align}
where $\mathbf{H}=[\mathbf{h}_1,\cdots,\mathbf{h}_S]\in \mathbb{C}^{M_{rx}\times S}$, and $\mathbf{g}_k\in\mathbb{C}^{N_{rx}\times 1}$ denote the channels from the SC BSs to the MC BS  and from the MC BS to the $k$-th SC BS, respectively.  $\mathbf{x}=[x_1,\cdots, x_S]^{T}\in \mathbb{C}^{S\times 1}$, $\mathbf{s}=[s_1,\cdots,s_S]^{T}\in \mathbb{C}^{S\times 1}$, $\mathbf{n}\in\mathbb{C}^{M_{rx}\times 1}$, and $\mathbf{v}_k\in\mathbb{C}^{N_{rx}\times 1}$ denote the transmit signal from the SC BSs to the MC BS,  from the MC BS to the SC BSs, the noise vectors at the MC BS, and the $k$-th SC BS, respectively. The elements of $\mathbf{x}$, $\mathbf{s}$, $\mathbf{n}$, and $\mathbf{v}_k$ are modeled by $\mathbb{CN}(0,1)$. $p_m$ and $p_s$ indicate the transmit power of the MC BS and the SC BSs, respectively.

$\mathbf{Q}\in\mathbb{C}^{M_{rx}\times M_{tx}}$, $\mathbf{q}_{s,k}\in \mathbb{C}^{N_{rx}\times 1}$, and $\mathbf{q}_{c,kj}\in \mathbb{C}^{N_{rx}\times 1}$ indicate the SI channel at the MC BS, the SI channel at the $k$-th SC BS, and the SC-to-SC interference channel from the $j$-th SC BS to the $k$-th SC BS, respectively.
Because the MC BS has perfect knowledge of its own transmitted signals, some form of active SI cancellation technique could be implemented, such that any residual interference emanating from the imperfect cancellation can be regarded as additional noise, with the same constraints as $\mathbf{s}$, i.e., $s_k\sim \mathbb{CN}(0,1), \forall k$. The residual loopback interference channel $\mathbf{Q}$ is modeled as Rayleigh \cite{Kong17}. Note that without hardware cancellation, we can efficiently suppress  the LoS components by antenna isolation and the major effects of SI originates from scattering. The common assumption is to model this residual interfering link as Rayleigh  \cite{Riihonen11}, \cite{Ngo14}, i.e., $\mathbf{Q}\sim \mathbb{CN}(0,\sigma^2_m\mathbf{I}_{M_{rx}})$. $\sigma^2_m$ can be interpreted as the SI strength, which is dependent on the distance separating the transmit and receive antenna arrays \cite{Ngo14} and/or hardware interference cancellation capability \cite{Riihonen2011}. Similar justification can be provided for the SI and the SC-to-SC interference at the SC BSs. Thus, we have $\mathbf{q}_{s,k}\sim\mathbb{CN}(\mathbf{0},\sigma^2_{s,k}\mathbf{I}_{N_{rx}})$, and $\mathbf{q}_{c,kj}\sim\mathbb{CN}(\mathbf{0},\sigma^2_{c,kj}\mathbf{I}_{N_{rx}})$.
The channels $\mathbf{H}$ and $\mathbf{g}_k$ are, respectively, modeled as \cite{Zhang14}
\begin{align}
\mathbf{H}=\mathbf{H}_L\sqrt{\pmb{\Theta}(\pmb{\Theta} + \mathbf{I}_{S})^{-1}} + \mathbf{H}_W\sqrt{(\pmb{\Theta} + \mathbf{I}_S)^{-1}},  
\end{align} 
%\vspace{-1.0cm}
\begin{align}
\mathbf{g}_k = \mathbf{g}_{L,k}\sqrt{K_{s,k}(K_{s,k} + 1)^{-1}} + \mathbf{g}_{W,k}\sqrt{(K_{s,k} + 1)^{-1}},
\end{align}
where $\mathbf{H}_L=\bar{\mathbf{H}}_L\mathbf{B}^{1/2}$, and $\mathbf{H}_W=\bar{\mathbf{H}}_W\mathbf{B}^{1/2}$ indicate the LoS deterministic and random Rayleigh components of $\mathbf{H}$, respectively.  $\mathbf{g}_{L,k}=\bar{\mathbf{g}}_{L,k}\alpha_k^{1/2}$ and $\mathbf{g}_{W,k}=\bar{\mathbf{g}}_{W,k}\alpha_k^{1/2}$ denote the LoS deterministic  and random Rayleigh fading components of $\mathbf{g}_k$, respectively. The elements of $\bar{\mathbf{H}}_W$ and $\bar{\mathbf{g}}_{W,k}$ are distributed by $\mathbb{CN}(0,1)$. The $S\times S$ diagonal matrices $\mathbf{B}$ and $\pmb{\Theta}$ contain the large-scale fading coefficients and Rician $K$-factors of $\mathbf{H}$, with the $k$-th elements $\beta_k$ and $K_{m,k}$, respectively. Also,  $\alpha_k$ and $K_{s,k}$  describe the large-scale fading coefficient and Rician $K$-factor of $\mathbf{g}_k$, respectively. The large-scale coefficients $\beta_k$ and $\alpha_k$ are assumed to stay constant over several coherence intervals and are known a priori. The LoS components $\bar{\mathbf{H}}_L$ (and $\bar{\mathbf{g}}_{L,k}$) are modeled as $[\bar{\mathbf{H}}_L]_{mk}=e^{-j(m-1)(2\pi d/\lambda)\sin(\theta_k)}$, where $d,\lambda$, and $\theta_k$ denote the antenna spacing, wavelength, and angle of arrival, respectively.\footnote{This modeling of the LoS components is restricted to uniform linear array (ULA) antennas \cite{Lota17}. However, it is generally employed in the massive MIMO literature since it yields tractable solutions and provide useful insights into the system design \cite{Wang17, Zhang2017, Zhang14}.}   Without loss of generality, we fix $d=\lambda/2$. 

The MC BS and $k$-th SC BS quantize the received signal before processing.  For tractability, we assume the additive quantization noise model (AQNM) for the receivers with low-resolution ADCs. The AQNM has been shown to be accurate at low to medium signal-to-noise ratios (SNRs) and it is a common assumption in the literature \cite{Kong17, Zhang2017, Zhang2019} and references therein.  The quantized signal at the MC BS and  $k$-th SC BS, are written, respectively, as
\begin{align}
\mathbf{y}^{(1)}_{qm} = \rho_1\mathbf{y}^{(1)}_m + \mathbf{n}_q\hspace{0.1cm},  
\end{align}
%
%\vspace{-1cm}
\begin{align}
\mathbf{y}^{(1)}_{qs,k} = \epsilon_1\mathbf{y}^{(1)}_{s,k} + \mathbf{v}_{q,k}\hspace{0.1cm},
\end{align}
where $\rho_1=1-\kappa$, and $\epsilon_1=1-\kappa$ describe the ADC resolution at the MC BS and the $k$-th SC BS, respectively. Approximate values of $\kappa$ are shown in the Table \ref{tablebits}\cite{Zhang2017}. For $b>5, \kappa=\frac{\pi\sqrt{3}}{2}\cdot 2^{-2b}$, where $b$ denotes the  quantization bits. $\mathbf{n}_q$ and $\mathbf{v}_{q,k}$ (uncorrelated with $\mathbf{y}_m$ and $\mathbf{y}_{s,k}$) indicate the additive Gaussian   QN vectors at the MC BS and $k$-th SC BS, which have the covariances $\mathbf{N}_q=\rho_1(1-\rho_1)\text{diag}(\mathbb{E}[\mathbf{y}_m^{(1)}\mathbf{y}_m^{(1)H}])$ and $\mathbf{V}_{q,k}=\epsilon_1(1-\epsilon_1)\text{diag}(\mathbb{E}[\mathbf{y}_{s,k}^{(1)}\mathbf{y}_{s,k}^{(1)H}])$, respectively. The MC BS and the $k$-th SC BS decode their signals using the filters $\mathbf{r}^{H}_{m,k}$ and $\mathbf{r}^{H}_{s,k}$ with the outputs represented, respectively, as 
\begin{align}\label{eqrecquanmc1}
y^{(1)}_{qm,k}& =\rho_1\sqrt{p_s}\mathbf{r}^{H}_{m,k}\mathbf{h}_{k}x_k + \rho_1\sqrt{p_s}\sum^S_{j=1,j\neq k}\mathbf{r}^{H}_{m,k}\mathbf{h}_{j}x_j\nonumber  \\&+ \rho_1\sqrt{p_m}\mathbf{r}^{H}_{m,k}\mathbf{Qs} + \rho_1\mathbf{r}^{H}_{m,k}\mathbf{n} + \mathbf{r}^{H}_{m,k}\mathbf{n}_q\hspace{0.1cm},
\end{align}  
%
%\vspace{-0.6cm}
\begin{align}\label{eqrecquansc1}
y^{(1)}_{qs,k} = \mathbf{r}^{H}_{s,k}\mathbf{y}^{(1)}_{qs,k}\hspace{0.1cm}.
\end{align}
\begin{table}[!t]
% increase table row spacing, adjust to taste
\renewcommand{\arraystretch}{1.0}%\vspace{-0.4cm}
% if using array.sty, it might be a good idea to tweak the value of
% \extrarowheight as needed to properly center the text within the cells
\caption{$\kappa$ for different $b$-bit ADC resolution }
\label{tablebits}
\centering
% Some packages, such as MDW tools, offer better commands for making tables
% than the plain LaTeX2e tabular which is used here.
\begin{tabular}{|c|c|c|c|c|c|}
\hline
$b$ &1 & 2 & 3 & 4 & 5 \\
\hline
$\kappa$ & 0.3634 &0.1175 &0.03454 &0.009497 &0.002499\\
\hline
\end{tabular}%\vspace{-0.3cm}
\end{table}
%
%
%\vspace{-1cm}
\subsection{Second Phase Data Communication}
Here, the MC BS precodes its signal and transmits to the $S$ SC BSs. The $k$-th SC BS precodes its data and sends the signal to the corresponding $k$-th receive antenna of the MC BS, simultaneously. 
%\footnote{For similar reason as the first phase, no form of receive beamformer is used at the MC BS during the second phase.} 
The received signal at the $k$-th MC BS  receive antenna and the $k$-th SC BS are, respectively, given as
\begin{align}
y^{(2)}_{m,k} &= \sqrt{p_s}\mathbf{g}_k^H\mathbf{f}_{s,k}x_k + \sqrt{p_s}\sum^S_{j=1,j\neq k}\mathbf{g}^H_k\mathbf{f}_{s,j}x_j\nonumber \\&+  \sqrt{p_m}\mathbf{z}_{m,k}\mathbf{F}_m\mathbf{s} + n_k,
\end{align}
%
%\vspace{-0.8cm}
\begin{align}
y^{(2)}_{s,k}&=\sqrt{p_m}\mathbf{h}_k^H\mathbf{f}_{m,k}s_k + \sqrt{p_m}\sum^S_{j=1,j\neq k}\mathbf{h}_k^H\mathbf{f}_{m,j}s_j \nonumber\\&+ \sqrt{p_s}\mathbf{z}_{s,k}\mathbf{f}_{s,k}x_k +\sqrt{p_s}\sum^S_{j=1,j\neq k}\mathbf{z}_{c,kj}\mathbf{f}_{s,j}x_j + v_{k},
\end{align}
where $\mathbf{f}_{s,k}\in \mathbb{C}^{N_{tx}\times 1}$, $\mathbf{F}_m=[\mathbf{f}_{m,1},\cdots,\mathbf{f}_{m,S}]\in \mathbb{C}^{M_{tx}\times S}$, $n_k$, and $v_k$ indicate the $k$-th SC BS and MC BS precoders, noise at the $k$-th MC BS receive antenna, and the $k$-th SC BS, respectively. $\mathbf{z}_{m,k}\in \mathbb{C}^{1\times M_{tx}}$, $\mathbf{z}_{s,k}\in \mathbb{C}^{1\times N_{tx}}$, and $\mathbf{z}_{c,kj}\in \mathbb{C}^{1\times N_{tx}}$ denote the SI channel at the $k$-th MC BS receive antenna, SI channel at the $k$-th SC BS, and SC-to-SC interference channel from the $j$-th SC BS to the $k$-th SC BS, respectively. The elements of  $\mathbf{z}_{m,k}$, $\mathbf{z}_{s,k}$, and $\mathbf{z}_{c,kj}$ are modeled by $\mathbb{CN}(0,\zeta^2_{m,k})$, $\mathbb{CN}(0,\zeta^2_{s,k})$, and $\mathbb{CN}(0,\zeta^2_{c,kj})$, respectively. $n_k$ and $v_k$ are distributed as $\mathbb{CN}(0,1)$. The signals received at the $k$-th receive antenna of the MC BS and the $k$-th SC BS are quantized before processing, where the outputs are expressed, respectively, as 
\begin{align}\label{eqrecquanmc2}
y^{(2)}_{qm,k} = \rho_2 y^{(2)}_{m,k} + n_{q,k} \hspace{0.1cm},
\end{align}
%
%\vspace{-1cm}
\begin{align}\label{eqrecquansc2}
y^{(2)}_{qs,k} = \epsilon_2 y^{(2)}_{s,k} + v_{q,k} \hspace{0.1cm},
\end{align}
where $\rho_2(\textrm{and}\hspace{0.1cm}\epsilon_2)=1-\kappa$ describe the ADC resolution at the $k$-th  receive antenna of the MC BS and the $k$-th SC BS, respectively. $n_{q,k}$ and $v_{q,k}$ denote the additive Gaussian QN at the MC BS $k$-th receive antenna and the $k$-th SC BS, with the covariances $N_{q,k}=\rho_2(1-\rho_2)\mathbb{E}[|y_{m,k}^{(2)}|^2]$ and $V_{q,k}=\epsilon_2(1-\epsilon_2)\mathbb{E}[|y_{s,k}^{(2)}|^2]$, respectively.
%
%\vspace{-0.2cm}
\subsection{Channel Estimation}
%To detect the received signal in the first phase and precode the signal for transmission in the second phase, the BSs  estimate the channel. 
As it is customary with massive MIMO analysis, we assume channel estimation by pilot sounding. Here, the MC BS transmit antennas send pilot signals to their corresponding SC BS for the DL channel estimation. The SC BSs, in turn, transmit their pilot signals to the MC BS for the UL estimation. To avoid the pilot contamination, the pilot sequences must be mutually orthogonal, i.e., the pilot length $\tau_p\geq 2S$ (in symbols) \cite{Ngo14,Anokye18}.\footnote{The MC BS and SC BSs can transmit their pilot signals simultaneously as in \cite{Ngo14} or the MC BS transmit antennas remain silent while the SC BSs send their pilot signals and vice versa as employed in \cite{Kong17}. However, both approaches achieve the same performance. Note that during the channel estimation phase, the MC BS has massive receive antennas and a few transmit antennas $M_{tx}=S$  and the $k$-th SC BS possesses massive receive antennas and a single transmit antenna.} The pilot length $\tau_p$ is less than the coherence time $T$. Assuming linear MMSE, the channels $\mathbf{H}$ and $\mathbf{g}_k$ are decomposed as \cite{Wang17}
\begin{align}\label{eqcsimc}
\mathbf{H} = \hat{\mathbf{H}} + \mathbf{E}\hspace{0.1cm}, 
\end{align}
%
%\vspace{-1.4cm}
\begin{align}\label{eqcsisc}
\mathbf{g}_k = \hat{\mathbf{g}}_k + \mathbf{d}_k\hspace{0.1cm},
\end{align}
where $\hat{\mathbf{H}}=[\hat{\mathbf{h}}_1,\cdots,\hat{\mathbf{h}}_S]\sim \mathbb{CN}(\mathbb{E}[\mathbf{\hat{H}}],\hat{\mathbf{B}})$, and $\mathbf{E}=[\mathbf{e}_1,\cdots,\mathbf{e}_S]\sim \mathbb{CN}(\mathbf{0},\pmb{\Xi}-\hat{\mathbf{B}})$ denote the estimated and error matrices of $\mathbf{H}$, respectively, which are mutually independent. $\pmb{\Xi}$ and $\hat{\mathbf{B}}$  are diagonal matrices with the $k$-th element $\xi_k\overset\Delta{=}\frac{\beta_k}{(K_{m,k} + 1)}$ and   
$\hat{\beta}_k\overset\Delta{=}\frac{\beta_k\eta_k}{(K_{m,k} + 1)}$, respectively, and $\eta_k=\frac{\tau_p p_\tau \beta_k}{(1+\tau_p p_\tau\beta_k)}$. Note that the $k$-th element of the error covariance matrix is $\tilde{\beta}_k\overset\Delta{=}\frac{\beta_k}{(K_{m,k}+1)(1+\tau_pp_\tau\beta_k)}$. 
 $\hat{\mathbf{g}}_k$ and $\mathbf{d}_k$ denote the estimated and error vectors of $\mathbf{g}_k$,  which are independent vectors with the variances $\hat{\alpha}_k\overset\Delta{=}\frac{\alpha_k\varepsilon_k}{(K_{s,k} + 1)}$, and $\tilde{\alpha}_k\overset\Delta{=}\frac{\alpha_k}{(K_{s,k}+1)(1+\tau_pp_\tau\alpha_k)}$, respectively, with $\varepsilon_k=\frac{\tau_p p_\tau \alpha_k}{(1+\tau_p p_\tau\alpha_k)}$. $p_\tau$ denotes the pilot power. 
\footnote{We follow the channel estimation approach in \cite{Liang16} and \cite{JZhang16}, where a few set of high-resolution ADCs , say $c$ (used to perform channel estimation only), are first connected to the first set of $c$ receive antennas to estimate the channel coefficients $h_{k,1},\cdots, h_{k,c}$. Then, in a round robin fashion, an optimal ADC switch switches to the next $c$ receive antennas to acquire the channel estimates $h_{k,c+1},\cdots,h_{k,2c}$.  Therefore, the QN effects on the channel estimates can be ignored. Please refer to \cite{Liang16} for the details.}

\subsection{Data Detection and Signal Precoding}      
To detect and precode the signal, the MC BS and the SC BSs treat the estimated channel as the true channel. We assume the maximum ratio combining/transmission (MRC/MRT) for the first phase data detection and second phase signal precoding, respectively, due to its simplicity and the fact that processing can be done in a distributed manner. For the first phase, the receive filter at the MC BS and  $k$-th SC BS are, respectively, given as 
$\mathbf{r}^H_{m,k}=\hat{\mathbf{h}}_k^H$ and $\mathbf{r}^H_{s,k}=\hat{\mathbf{g}}_k^H$.
%\begin{align}\label{eq:detect}
%\mathbf{r}^H_{m,k}=\hat{\mathbf{h}}_k^H,\quad \mathbf{r}^H_{s,k}=\hat{\mathbf{g}}_k^H\hspace{0.1cm}.
%\end{align}
%
%
For the second phase, the MC BS and the $k$-th SC BS precoders are obtained as $\mathbf{F}_m=\mu_m\hat{\mathbf{H}}$ and $\mathbf{f}_{s,k}=\mu_{s,k}\hat{\mathbf{g}}_k$, respectively. $\mu_m$ and $\mu_{s,k}$ describe the power normalization factors defined by
\begin{align*}
&\mu_m^2=S/\textrm{tr}(\mathbb{E}[\hat{\mathbf{H}}\hat{\mathbf{H}}^H])=\dfrac{S}{M_{tx}\sum\limits^S_{j=1}\beta_j\frac{(K_{m,j}+\eta_j)}{(K_{m,j} + 1)}},\\&
\mu^2_{s,k}=1/\textrm{tr}(\mathbb{E}[\hat{\mathbf{g}}_k\hat{\mathbf{g}}_k^H])=\dfrac{1}{N_{tx}\alpha_k\frac{K_{s,k} + \varepsilon_k}{K_{s,k}+1}}.
\end{align*}
%The next section analyzes the UL/DL SE of the proposed HetNet topology.
%
%
\section{Backhaul Spectral Efficiency Analysis}\label{secrate}
In this section, we study the analytic expressions for the UL/DL SE of the HetNet model. For a massive MIMO system, the following achievable rate expression holds \cite{Zhang14}:
\begin{align}\label{eqrate}
\mathbb{E}\Bigg[\log_2\Bigg(1 + \frac{\mathbb{A}}{\mathbb{B}}\Bigg)\Bigg]\approx \log_2\Bigg(1 + \frac{\mathbb{E}[\mathbb{A}]}{\mathbb{E}[\mathbb{B}]}\Bigg),
\end{align}
where $\mathbb{A}$ and $\mathbb{B}$ denote the desired signal and the interference-plus-noise, respectively. Due to the intractability of the left hand side (LHS) of \eqref{eqrate}, we use the approximation on the right hand side (RHS). This has been proved to be tight in the massive antenna regime \cite{Zhang14}.
To derive the analytic SE  for the first phase, we rewrite \eqref{eqrecquanmc1} and \eqref{eqrecquansc1} based on \eqref{eqcsimc} and \eqref{eqcsisc}, respectively,  as  
\begin{align}\label{eqrecsemc1}
&y^{(1)}_{qm,k} =\underbrace{\rho_1\sqrt{p_s}\mathbf{r}^{H}_{m,k}\hat{\mathbf{h}}_{k}x_k}_{\text{Desired Signal}} + \underbrace{\rho_1\sqrt{p_s}\sum^S_{j\neq k}\mathbf{r}^{H}_{m,k}\hat{\mathbf{h}}_{j}x_j}_{\text{Intercell Interference}} + \underbrace{\rho_1\mathbf{r}^{H}_{m,k}\mathbf{n}}_{\text{Noise}} \nonumber\\& 
+ \underbrace{\rho_1\sqrt{p_s}\sum^S_{j=1}\mathbf{r}^{H}_{m,k}\mathbf{e}_{j}x_j}_{\text{Channel Estimation Error}}  + \underbrace{\rho_1\sqrt{p_m}\mathbf{r}^{H}_{m,k}\mathbf{Qs}}_{\text{SI}}  + \underbrace{\mathbf{r}^{H}_{m,k}\mathbf{n}_q}_{\text{QN}},
\end{align} 
%and
%\vspace{-0.7cm}
\begin{align}\label{eqrecsesc1}
&y^{(1)}_{qs,k} =\underbrace{\epsilon_1\sqrt{p_m}\mathbf{r}^{H}_{s,k}\hat{\mathbf{g}}_{k}s_k}_{\text{Desired Signal}} + \underbrace{\epsilon_1\sqrt{p_m}\sum^S_{j\neq k}\mathbf{r}^{H}_{s,k}\hat{\mathbf{g}}_{j}s_j}_{\text{Intercell Interference}}\nonumber\\&
+ \underbrace{\epsilon_1\sqrt{p_m}\sum^S_{j=1}\mathbf{r}^{H}_{s,k}\mathbf{d}_{j}s_j}_{\text{Channel Estimation Error}}  + \underbrace{\epsilon_1\sqrt{p_s}\mathbf{r}^{H}_{m,k}\mathbf{q}_{s,k}x_k}_{\text{SI}} \nonumber\\& + \underbrace{\epsilon_1\sqrt{p_s}\sum^S_{j\neq k}\mathbf{r}_{s,k}^H\mathbf{q}_{c,kj}x_j}_{\text{SC-to-SC Interference}} +\underbrace{\epsilon_1\mathbf{r}^{H}_{s,k}\mathbf{v}_k}_{\text{Noise}}  + \underbrace{\mathbf{r}^{H}_{s,k}\mathbf{v}_{q,k}}_{\text{QN}}\hspace{0.1cm}.
\end{align} 
By using \eqref{eqrate}, \eqref{eqrecsemc1}, and \eqref{eqrecsesc1},  the UL/DL SE for the first phase, are determined, respectively, as
\begin{align}\label{eqratemc1}
&R^{(1)}_{m,k} = \tau_d\log_2\Bigg(1 + \dfrac{\mathbb{E}[\mathbb{A}^{(1)}_{m,k}]}{\mathbb{E}[\mathbb{B}^{(1)}_{m,k}] + \mathbb{E}[\mathbb{D}^{(1)}_{m,k}]}\Bigg), \hspace{0.1cm} \textrm{and}\nonumber\\& \hspace{0.1cm} R^{(1)}_{s,k} = \tau_d\log_2\Bigg(1 +\dfrac{\mathbb{E}[\mathbb{A}^{(1)}_{s,k}]}{\mathbb{E}[\mathbb{B}^{(1)}_{s,k}] + \mathbb{E}[\mathbb{D}^{(1)}_{s,k}]}\Bigg),
\end{align}
%\vspace{-0.2cm}
%%and
%\begin{align}\label{eq:rate_sc1}\vspace{-0.3cm}
%R^{(1)}_{s,k} = \tau_d\log_2\Bigg(1 +\dfrac{\mathbb{A}^{(1)}_{s,k}}{\mathbb{B}^{(1)}_{s,k} + \mathbb{C}^{(1)}_{s,k}}\Bigg),
%\end{align}
%%
where $\tau_d=(T-\tau_p)/2T$, $\mathbb{A}^{(1)}_{m,k}=\rho_1^2p_s\mathbb{E}\big[\|\hat{\mathbf{h}}_k\|^4]$, $\mathbb{D}^{(1)}_{m,k}=\mathbb{E}\big[\|\hat{\mathbf{h}}^H_k\mathbf{n}_q\|^2 ]$, and
\newline
$\mathbb{B}^{(1)}_{m,k}=\rho_1^2p_s\sum^S_{j\neq k}\mathbb{E}[|\hat{\mathbf{h}}^H_k\hat{\mathbf{h}}_j|^2] + \rho_1^2p_s\sum^S_{j=1}\mathbb{E}[|\hat{\mathbf{h}}^H_k\mathbf{e}_j|^2]+\rho_1^2p_m\mathbb{E}[|\hat{\mathbf{h}}^H_k\mathbf{Q}|^2] + \rho_1^2\mathbb{E}\big[\|\hat{\mathbf{h}}^H_k\|^2]$.
%
%\begin{align*}
%\mathbb{A}^{(1)}_{m,k}=\rho_1^2p_s\mathbb{E}\big[\|\hat{\mathbf{h}}_k\|^4],\quad \mathbb{C}^{(1)}_{m,k}=\mathbb{E}\big[\|\hat{\mathbf{h}}^H_k\mathbf{n}_q\|^2 ],\hspace{0.1cm}\textrm{and}
%\end{align*}
%
%%\vspace{-1cm}
%\begin{align*}
%\mathbb{B}^{(1)}_{m,k}=&\rho_1^2p_s\sum^S_{j\neq k}\mathbb{E}[|\hat{\mathbf{h}}^H_k\hat{\mathbf{h}}_j|^2] + \rho_1^2p_s\sum^S_{j=1}\mathbb{E}[|\hat{\mathbf{h}}^H_k\mathbf{e}_j|^2]+\rho_1^2p_m\mathbb{E}[|\hat{\mathbf{h}}^H_k\mathbf{Q}|^2] + \rho_1^2\mathbb{E}\big[\|\hat{\mathbf{h}}^H_k\|^2].
%\end{align*}
Also, we have $\mathbb{A}^{(1)}_{s,k}=\epsilon_1^2p_m\mathbb{E}\big[\|\hat{\mathbf{g}}_k\|^4],\quad  \mathbb{D}^{(1)}_{s,k}=\mathbb{E}\big[\|\hat{\mathbf{g}}^H_k\mathbf{v}_q\|^2 ]$, and $\mathbb{B}^{(1)}_{s,k}=\epsilon_1^2p_m\sum^S_{j\neq k}\mathbb{E}[|\hat{\mathbf{g}}^H_k\hat{\mathbf{g}}_j|^2] + \epsilon_1^2p_m\sum^S_{j=1}\mathbb{E}[|\hat{\mathbf{g}}^H_k\mathbf{d}_j|^2]+\epsilon_1^2p_s\mathbb{E}[|\hat{\mathbf{g}}^H_k\mathbf{q}_{s,k}|^2]+ \epsilon_1^2p_s\sum\limits^S_{j\neq k}\mathbb{E}[|\hat{\mathbf{g}}^H_k\hat{\mathbf{q}}_{c,kj}|^2] + \epsilon_1^2\mathbb{E}\big[\|\hat{\mathbf{g}}^H_k\|^2]$.
%
%\begin{align*}
%\mathbb{A}^{(1)}_{s,k}=\epsilon_1^2p_m\mathbb{E}\big[\|\hat{\mathbf{g}}_k\|^4],\quad  \mathbb{C}^{(1)}_{s,k}=\mathbb{E}\big[\|\hat{\mathbf{g}}^H_k\mathbf{v}_q\|^2 ], \hspace{0.1cm}\textrm{and}
%\end{align*}
%
%\vspace{-1.0cm}
%\begin{align*}
%&\mathbb{B}^{(1)}_{s,k}=\epsilon_1^2p_m\sum^S_{j\neq k}\mathbb{E}[|\hat{\mathbf{g}}^H_k\hat{\mathbf{g}}_j|^2] + \epsilon_1^2p_m\sum^S_{j=1}\mathbb{E}[|\hat{\mathbf{g}}^H_k\mathbf{d}_j|^2]+\epsilon_1^2p_s\mathbb{E}[|\hat{\mathbf{g}}^H_k\mathbf{q}_{s,k}|^2]+ \epsilon_1^2p_s\sum\limits^S_{j\neq k}\mathbb{E}[|\hat{\mathbf{g}}^H_k\hat{\mathbf{q}}_{c,kj}|^2] \\&+ \epsilon_1^2\mathbb{E}\big[\|\hat{\mathbf{g}}^H_k\|^2].
%\end{align*}
%
The terms $\mathbb{A}^{(1)}_{m,k}$ and $\mathbb{D}^{(1)}_{m,k}$ describe the desired signal at the MC BS for the $k$-th SC BS and the impact of the QN, respectively, while $\mathbb{B}^{(1)}_{m,k}$ accounts for the intercell interference (ICI), channel estimation error, SI, and noise. $\mathbb{A}^{(1)}_{s,k}$ and $\mathbb{D}^{(1)}_{s,k}$ define the desired signal and the QN at the $k$-th SC BS, respectively. Moreover, $\mathbb{B}^{(1)}_{s,k}$ considers the ICI, channel estimation error, SI, SC-to-SC interference, and noise.  

Similarly, for the second phase, we  rewrite \eqref{eqrecquanmc2} and \eqref{eqrecquansc2}, respectively, as 
\begin{align}\label{eqrecsemc2}
&y^{(2)}_{qm,k} = \underbrace{\rho_2\sqrt{p_s}\hat{\mathbf{g}}_k^H\mathbf{f}_{s,k}x_k}_{\text{Desired Signal}} + \underbrace{\rho_2\sqrt{p_s}\sum^S_{j=1,j\neq k}\hat{\mathbf{g}}^H_k\mathbf{f}_{s,j}x_j}_{\text{Intercell Interference}}+ \underbrace{\rho_2n_k}_{\text{Noise}}\nonumber\\&  + \underbrace{\rho_2\sqrt{p_s}\sum^S_{j=1}\mathbf{d}^H_k\mathbf{f}_{s,j}x_j}_{\text{Channel Estimation Error}}+ \underbrace{\rho_2\sqrt{p_m}\mathbf{z}_{m,k}\mathbf{F}_m\mathbf{s}}_{\text{SI}} + \underbrace{n_{q,k}}_{\text{QN}},
\end{align}
%
%\vspace{-0.7cm}
\begin{align}\label{eqrecsesc2}
&y^{(2)}_{qs,k}=\underbrace{\epsilon_2\sqrt{p_m}\hat{\mathbf{h}}_k^H\mathbf{f}_{m,k}s_k}_{\text{Desired Signal}} + \underbrace{\epsilon_2\sqrt{p_m}\sum^S_{j=1,j\neq k}\hat{\mathbf{h}}_k^H\mathbf{f}_{m,j}s_j}_{\text{Intercell Interference}} \nonumber\\&+ \underbrace{\epsilon_2\sqrt{p_m}\sum^S_{j=1}\mathbf{e}_k^H\mathbf{f}_{m,j}s_j}_{\text{Channel Estimation Error}} + \underbrace{\epsilon_2\sqrt{p_s}\mathbf{z}_{s,k}\mathbf{f}_{s,k}x_k}_{\text{SI}} \nonumber\\&+\underbrace{\epsilon_2\sqrt{p_s}\sum^S_{j=1,j\neq k}\mathbf{z}_{c,kj}\mathbf{f}_{s,j}s_j}_{\text{SC-to-SC Interference}} + \underbrace{\epsilon_2v_{k}}_{\text{Noise}} + \underbrace{v_{q,k}}_{\text{QN}}\hspace{0.1cm}.
\end{align}
Using \eqref{eqrecsemc2} and \eqref{eqrecsesc2}, the UL/DL SE at the MC BS and the $k$-th SC BS, for the second phase, are respectively, given by
\begin{align}\label{eqratemc2}
&R^{(2)}_{m,k} = \tau_d\log_2\Bigg(1 + \dfrac{\mathbb{E}[\mathbb{A}^{(2)}_{m,k}]}{\mathbb{E}[\mathbb{B}^{(2)}_{m,k}] +  \mathbb{E}[N_{q,k}]}\Bigg),\hspace{0.1cm} \textrm{and}\nonumber\\& R^{(2)}_{s,k} = \tau_d\log_2\Bigg(1 + \dfrac{\mathbb{E}[\mathbb{A}^{(2)}_{s,k}]}{\mathbb{E}[\mathbb{B}^{(2)}_{s,k}]+ \mathbb{E}[V_{q,k}]}\Bigg),
\end{align}
where $\mathbb{A}^{(2)}_{m,k} = \rho_2^2\mu_{s,k}^2p_s\mathbb{E}\big[\|\hat{\mathbf{g}}_k\|^4]$ and $\mathbb{B}^{(2)}_{m,k}=\rho_2^2p_s\sum^S_{j\neq k}\mu^2_{s,j} \mathbb{E}[|\hat{\mathbf{g}}^H_k\hat{\mathbf{g}}_j|^2] + \rho_2^2p_s\sum^S_{j=1}\mu^2_{s,j}\times \\\mathbb{E}[|\mathbf{d}_k^H\hat{\mathbf{g}}_j|^2]+\rho_2^2\mu^2_{m}p_m\mathbb{E}[|\mathbf{z}_{m,k}\hat{\mathbf{H}}|^2] + \rho_2^2$.
We also define $\mathbb{A}^{(2)}_{s,k} = \epsilon_2^2\mu_m^2p_m\mathbb{E}\big[\|\hat{\mathbf{h}}_k\|^4 ]$ and $\mathbb{B}^{(2)}_{s,k}=\epsilon_2^2p_m\sum\limits^S_{j\neq k}\mu_m^2\mathbb{E}[|\hat{\mathbf{h}}^H_k\hat{\mathbf{h}}_j|^2] +\epsilon_2^2p_m\sum\limits^S_{j=1} \mu_m^2\mathbb{E}[|\mathbf{e}_k^H\hat{\mathbf{h}}_j|^2]+\epsilon_2^2\mu_{s,k}^2p_s\mathbb{E}[|\mathbf{z}_{s,k}\hat{\mathbf{g}}_k|^2]+ \epsilon_2^2p_s\sum\limits^S_{j\neq k}\mu_{s,j}^2\mathbb{E}[|\mathbf{z}_{c,kj}\hat{\mathbf{g}}_j|^2] + \epsilon_2^2$.  

The overall SE of the system is affected by several factors. To appreciate the influence of the various parameters,
in the next section, we derive closed-form solutions for the UL/DL SE.

\section{Large System Analysis}\label{seclargesysanalysis}
To obtain closed-form solutions for the SE of the first phase, we assume the following: The receive antennas at the MC BS $M_{rx}\rightarrow\infty$ with the transmit antennas fixed as $M_{tx}=S$, and receive antennas at the $k$-th SC BS $N_{rx}\rightarrow \infty$ with a single transmit antenna. For the second phase, due to the switching of the roles of the antennas, we allow the transmit antennas at the MC BS $M_{tx}\rightarrow \infty$ while the receive antennas are fixed such that $M_{rx}=S$. For the $k$-th SC BS, the transmit antennas $N_{tx}\rightarrow \infty$ with a single receive antenna. By using the law of large numbers and some results from \cite{Zhang14}, the UL/DL SE for the first and second phases are obtained by Lemma 1 and Lemma 2, respectively.

\textit{Lemma 1}: By adopting the MRC filter and assuming  imperfect CSI, the UL/DL SE of the low-resolution ADC quantized FD massive MIMO-enabled backhaul of the proposed HetNet over  Rician channels, for the first phase, are obtained by \eqref{eqrateanalmc1} and \eqref{eqrateanalsc1}, respectively.
\begin{align}\label{eqrateanalmc1}
R^{(1)}_{m,k} = \tau_d\log_2\Bigg(1 +   \dfrac{\tilde{\mathbb{A}}_{m,k}^{(1)}}{\tilde{\mathbb{B}}^{(1)}_{m,k} + \tilde{\mathbb{D}}_{m,k}^{(1)}}\Bigg),
\end{align}
where $\tilde{\mathbb{A}}_{m,k}^{(1)}$, $\tilde{\mathbb{B}}^{(1)}_{m,k}$, and $\tilde{\mathbb{D}}_{m,k}^{(1)}$ are given by 
\begin{align*}
&\tilde{\mathbb{A}}_{m,k}^{(1)}=\rho_1p_s\beta_kM_{rx}[M_{rx}K^2_{m,k} +\eta_k(1+M_{rx})(2K_{m,k} + \eta_k)],\\&
\tilde{\mathbb{B}}^{(1)}_{m,k}=\rho_1(K_{m,k}+1)\Big[(K_{m,k}+\eta_k)M_{rx}\delta^{(1)}_{m,k}\\&+p_s\sum^S_{j\neq k}\frac{\beta_j}{K_{m,j}+1}\Delta^{(1)}_{m,kj}\Big],\\&
\tilde{\mathbb{D}}^{(1)}_{m,k}=(1-\rho_1)(K_{m,k}+1)M_{rx}\Big[(K_{m,k} +\eta_k)\delta^{(1)}_{m,k} \\&+\frac{p_s\beta_k}{K_{m,k}+1}(K_{m,k}^2+4K_{m,k}\eta_k + 2\eta_k^2) + p_s\sum^S_{j\neq k}\frac{\beta_j}{K_{m,j}+1}\times\\&(K_{m,j}\eta_k + K_{m,k}\eta_j +\eta_k\eta_j)\Big],
\end{align*}
with $\delta^{(1)}_{m,k}\overset\Delta{=}p_s\sum^S_{j=1}\tilde{\beta}_j+p_m\sigma^2_{m}S+1$,
%\newline
$\Delta^{(1)}_{m,kj}\overset\Delta{=}K_{m,k}K_{m,j}\phi_{m,kj}^2 + M_{rx}(K_{m,j}\eta_k +K_{m,k}\eta_j +\eta_k\eta_j)$,
and $\phi_{m,kj}^2\overset\Delta{=}\dfrac{\sin(\frac{M_{rx}\pi}{2}[\sin(\varphi_{m,k})-\sin(\varphi_{m,j})])}{\sin(\frac{\pi}{2}[\sin(\varphi_{m,k})-\sin(\varphi_{m,j})])}$.
\begin{align}\label{eqrateanalsc1}
R^{(1)}_{s,k} = \tau_d\log_2\Bigg(1 +   \dfrac{\tilde{\mathbb{A}}_{s,k}^{(1)}}{\tilde{\mathbb{B}}^{(1)}_{s,k} + \tilde{\mathbb{D}}_{s,k}^{(1)}}\Bigg),
\end{align}
where $\tilde{\mathbb{A}}_{s,k}^{(1)}$, $\tilde{\mathbb{B}}^{(1)}_{s,k}$, and $\tilde{\mathbb{D}}_{s,k}^{(1)}$ are defined as
\begin{align*}
&\tilde{\mathbb{A}}_{s,k}^{(1)}=\epsilon_1p_s\alpha_kN_{rx}[N_{rx}K^2_{s,k} +\varepsilon_k(1+N_{rx})(2K_{s,k} + \varepsilon_k)],\\&
\tilde{\mathbb{B}}^{(1)}_{s,k}=\epsilon_1(K_{s,k}+1)[(K_{s,k}+\varepsilon_k)N_{rx}\delta^{(1)}_{s,k}+p_m\sum^S_{j\neq k}\frac{\alpha_j\Delta^{(1)}_{s,kj}}{K_{s,j}+1}],\\&
\tilde{\mathbb{D}}^{(1)}_{s,k}=(1-\epsilon_1)(K_{s,k}+1)N_{rx}[(K_{s,k} +\varepsilon_k)\delta^{(1)}_{s,k}\\& +\frac{p_m\alpha_k}{K_{s,k}+1}(K_{s,k}^2+4K_{s,k}\varepsilon_k + 2\varepsilon_k^2) \\&+ p_m\sum^S_{j\neq k}\frac{\alpha_j}{K_{s,j}+1}(K_{s,j}\varepsilon_k + K_{s,k}\varepsilon_j +\varepsilon_k\varepsilon_j)],
\end{align*}
with $\delta^{(1)}_{s,k}\overset\Delta{=}p_m\sum^S_{j=1}\tilde{\alpha}_j+p_s\sigma^2_{s,k} + \sum^S_{j\neq k}p_s\sigma^2_{c,kj} +1$, 
%\newline
$\Delta^{(1)}_{s,k}\overset\Delta{=}K_{s,k}K_{s,j}\phi_{s,kj}^2 + N_{rx}(K_{s,j}\varepsilon_k +K_{s,k}\varepsilon_j +\varepsilon_k\varepsilon_j)$, and
$\phi_{s,kj}^2\overset\Delta{=}\dfrac{\sin(\frac{N_{rx}\pi}{2}[\sin(\varphi_{s,k})-\sin(\varphi_{s,j})])}{\sin(\frac{\pi}{2}[\sin(\varphi_{s,k})-\sin(\varphi_{s,j})])}$.
Also, $\varphi_{m,k}$ and $\varphi_{s,k}$ describe the angle of arrival at the MC BS and the $k$-th SC BS, respectively.

\textit{Proof}: Please refer to Appendix A.

From \eqref{eqrateanalmc1}, we notice that the desired signal at the MC BS $\tilde{\mathbb{A}}_{m,k}^{(1)}$ monotonically increases with the receive antennas $M_{rx}$. However, the rate is constrained by the ICI, channel estimation errors, and SI through the term $\tilde{\mathbb{B}}_{m,k}^{(1)}$. It can be seen via $\delta_{m,k}^{(1)}$ that the SI and channel estimation errors at the MC BS are enhanced by the number of SC BSs; the ICI also grows with the  SC BSs.   Furthermore,  $\tilde{\mathbb{D}}_{m,k}^{(1)}$ describes the QN whose influence grows as a function of the received signal strength. As the received signal power increases, the QN effect  grows. However, the desired signal has a higher order than the QN and thus the performance loss due to QN can be compensated by increasing the receive antennas $M_{rx}$. At the $k$-th SC BS, (i.e., according to \eqref{eqrateanalsc1}),  the desired signal  $\tilde{\mathbb{A}}_{s,k}^{(1)}$ increases with the receive antennas $N_{rx}$. The SE at the $k$-th SC BS is affected by the channel estimation error, ICI, SI, SC-to-SC interference (through the term $\tilde{\mathbb{B}}_{s,k}^{(1)}$),  and the QN $\tilde{\mathbb{D}}_{s,k}^{(1)}$. It is observed through $\delta_{s,k}^{(1)}$ that the SC-to-SC interference and estimation errors increase with the number of SC BSs. As $\rho_1$ (and $\epsilon_1\rightarrow 1$), the QN terms $\tilde{\mathbb{D}}_{m,k}^{(1)}$ and  $\tilde{\mathbb{D}}_{s,k}^{(1)}$ tend to zero --- the SEs approach the ideal ADC case.

\textit{Lemma 2}: Assuming  an imperfect knowledge of the CSI in Rician fading channels, and utilizing the MRT precoding in the second phase, the UL/DL SEs of the low-resolution ADC quantized FD massive MIMO-enabled backhaul are approximated by \eqref{eqrateanalmc2} and \eqref{eqrateanalsc2}, respectively.   
%\vspace{-0.1cm}
\begin{align}\label{eqrateanalmc2}
R^{(2)}_{m,k} = \tau_d\log_2\Bigg(1 +   \dfrac{\rho_2\tilde{\mathbb{A}}_{m,k}^{(2)}}{\rho_2\tilde{\mathbb{B}}^{(2)}_{m,k} + \tilde{N}_{q,k}}\Bigg),
\end{align}
where $\tilde{N}_{q,k} =(1-\rho_2)(\tilde{\mathbb{A}}^{(2)}_{m,k} + \tilde{\mathbb{B}}^{(2)}_{m,k})$,
\begin{align*}
&\tilde{\mathbb{A}}_{m,k}^{(2)}=p_s\frac{\alpha_k[N_{tx}K_{s,k}^2+\varepsilon_k(1+N_{tx})(2K_{s,k}+\varepsilon_k)]}{(K_{s,k} + \varepsilon_k)(K_{s,k}+1)},\\& \tilde{\mathbb{B}}^{(2)}_{m,k}= \delta^{(2)}_{m,k} + \sum^S_{j\neq k}\frac{p_s\alpha_k\Delta^{(2)}_{m,kj}}{N_{tx}(K_{s,k}+1)(K_{s,j}+\varepsilon_j)},
\end{align*}
with $\delta^{(2)}_{m,k}\overset\Delta{=} p_m\zeta_{m,k}^2S + p_sS\tilde{\alpha}_k^2 +1$, 
\newline
$\Delta_{m,kj}^{(2)}\overset\Delta{=} K_{s,k}K_{s,j}\psi_{m,kj}^2 + N_{tx}(K_{s,j}\varepsilon_k + K_{s,k}\varepsilon_j + \varepsilon_j\varepsilon_k)$, and
\newline
$\psi_{m,kj}^2\overset\Delta{=}\dfrac{\sin(\frac{N_{tx}\pi}{2}[\sin(\theta_{m,k})-\sin(\theta_{m,j})])}{\sin(\frac{\pi}{2}[\sin(\theta_{m,k})-\sin(\theta_{m,j})])}$.
%
%\begin{align*}
%\psi_{m,kj}^2\overset\Delta{=}\dfrac{\sin(\frac{N_{tx}\pi}{2}[\sin(\theta_{m,k})-\sin(\theta_{m,j})])}{\sin(\frac{\pi}{2}[\sin(\theta_{m,k})-\sin(\theta_{m,j})])}.
%\end{align*}
%
\begin{align}\label{eqrateanalsc2}
R^{(2)}_{s,k} = \tau_d\log_2\Bigg(1 +   \dfrac{\epsilon_2\tilde{\mathbb{A}}_{s,k}^{(2)}}{\epsilon_2\tilde{\mathbb{B}}^{(2)}_{s,k} + \tilde{V}_{q,k}^{(2)}}\Bigg),
\end{align}
where $\tilde{V}_{q,k}^{(2)}= (1-\epsilon_2)(\tilde{\mathbb{A}}_{s,k}^{(2)} + \tilde{\mathbb{B}}_{s,k}^{(2)})$,
\begin{align*}
&\tilde{\mathbb{A}}_{s,k}^{(2)}=\dfrac{p_m\beta^2_kS[M_{tx}K_{m,k}^2 + \eta_k(1+M_{tx})(2K_{m,k} +\eta_k)]}{(K_{m,k} + 1)^2\sum^S_{j=1}\beta_j(K_{m,j}+\eta_j)(K_{m,j}+1)^{-1}},\\& \tilde{\mathbb{B}}_{s,k}^{(2)}=\delta_{s,k}^{(s)} + \sum^S_{j\neq k}\frac{p_m\mu_m^2\beta_k\beta_j\Delta_{s,kj}^{(2)}}{(K_{m,k}+1)(K_{m,j}+ 1)}.
\end{align*}
%
%\begin{align*}
%\tilde{\mathbb{B}}_{s,k}^{(2)}=\delta_{s,k}^{(s)} + p_m\sum^S_{j\neq k}\mu_m^2\frac{\beta_k\beta_j}{(K_{m,k}+1)(K_{m,j}+ 1)}\Delta_{s,kj}^{(2)},
%\end{align*}
%\begin{align*}
%\tilde{V}_{q,k}^{(2)}= (1-\epsilon_2)(\tilde{\mathbb{A}}_{s,k}^{(2)} + \tilde{\mathbb{B}}_{s,k}^{(2)}).
%\end{align*}
We also define $\delta_{s,k}^{(2)}\overset\Delta{=}p_s\zeta^2_{s,k} + p_s\sum^S_{j\neq k}\zeta_{c,kj}^2 + p_mS\tilde{\beta}_k^2 + 1$, $\Delta_{s,kj}^{(2)}\overset\Delta{=}K_{m,k}K_{m,j}\psi^2_{s,kj}+ M_{tx}(K_{m,j}\eta_k + K_{m,k}\eta_j + \eta_k\eta_j)$, and $\psi_{s,kj}^2\overset\Delta{=}\dfrac{\sin(\frac{M_{tx}\pi}{2}[\sin(\theta_{s,k})-\sin(\theta_{s,j})])}{\sin(\frac{\pi}{2}[\sin(\theta_{s,k})-\sin(\theta_{s,j})])}$.
%
%\begin{align*}
%\psi_{s,kj}^2\overset\Delta{=}\dfrac{\sin(\frac{M_{tx}\pi}{2}[\sin(\theta_{s,k})-\sin(\theta_{s,j})])}{\sin(\frac{\pi}{2}[\sin(\theta_{s,k})-\sin(\theta_{s,j})])}.
%\end{align*}
%%
$\theta_{m,k}$ and $\theta_{s,k}$ denote the angle of arrival at the $k$-th MC BS receive antenna and the $k$-th SC BS, respectively.

\textit{Proof}: Please follow the steps in the Appendix A.

Here, we observe that the SI at the MC BS and the SC-to-SC interference at the $k$-th SC BS are exacerbated by the number of SC BSs as shown in \eqref{eqrateanalmc2} (i.e., through the term $\delta^{(2)}_{m,k}$) and \eqref{eqrateanalsc2} (i.e., via the term $\delta^{(2)}_{s,k}$). The channel estimation error of the $k$-th SC BS also increases with the number of SC BSs. The analysis shows that the desired signal and the QN have equal order, signifying that the massive transmit antennas may not effectively suppress the QN.

\newcounter{MYtempeqncnt}
\begin{figure*}[!t]
% ensure that we have normalsize text
\normalsize
% Store the current equation number.
\setcounter{MYtempeqncnt}{\value{equation}}
% Set the equation number to one less than the one
% desired for the first equation here.
% The value here will have to changed if equations
% are added or removed prior to the place these
% equations are referenced in the main text.
\setcounter{equation}{25}
\begin{equation}
\label{eqrateconvergemc1}
\tilde{R}^{(1)}_{m,k} = \tau_d\log_2\Bigg(1 +   \dfrac{\rho_1p_s\beta_kM_{rx}}{\rho_1(p_m\sigma^2_mS + \frac{p_s}{M_{rx}}\sum^S_{j\neq k} \beta_j\phi_{m,kj}^2+ 1 )+ (1-\rho_1)(p_m\sigma^2_mS + p_s\beta_k + 1)} \Bigg),
\end{equation}
%%%
\begin{equation}
\label{eqrateconvergesc1}
\tilde{R}^{(1)}_{s,k} = \tau_d\log_2\Bigg(1 +  \dfrac{\epsilon_1p_m\alpha_kN_{rx}}{\epsilon_1(\sigma^2_{t,k} + \frac{p_m}{N_{rx}}\sum\limits^S_{j\neq k} \alpha_j\phi_{s,kj}^2+ 1 )+ (1-\epsilon_1)(\sigma^2_{t,k} +p_m\alpha_k + 1)} \Bigg),
\end{equation}
%%%
% Restore the current equation number.
\setcounter{equation}{\value{MYtempeqncnt}}
% IEEE uses as a separator
\hrulefill
% The spacer can be tweaked to stop underfull vboxes.
%\vspace*{2pt}
\end{figure*}

\subsection{Performance Evaluation}
Here, we evaluate the system  performance with respect to the Rician $K$-factor, the QN, the massive receive antennas (in the first phase), and massive transmit antennas (in the second phase).

\textit{Remark 1}: Let $K_{m,k}=K_m, \hspace{0.1cm}\forall k$;  $K_{s,k}=K_s, \forall k$, and $K_m=K_s=K$. Then as $K\rightarrow \infty$, the SE at the MC BS and the $k$-th SC BS, for the first phase,  converge to \eqref{eqrateconvergemc1} and \eqref{eqrateconvergesc1}, respectively (shown on top of this page), 
%%
%\begin{equation}
%\label{eq:rate_converge_mc1}
%\tilde{R}^{(1)}_{m,k} = \tau_d\log_2\Bigg(1 +   \dfrac{\rho_1p_s\beta_kM_{rx}}{\rho_1(p_m\sigma^2_mS + \frac{p_s}{M_{rx}}\sum^S_{j\neq k} \beta_j\phi_{m,kj}^2+ 1 )+ (1-\rho_1)(p_m\sigma^2_mS + p_s\beta_k + 1)} \Bigg),
%\end{equation}
%%%%
%\begin{equation}
%\label{eq:rate_converge_sc1}
%\tilde{R}^{(1)}_{s,k} = \tau_d\log_2\Bigg(1 +  \dfrac{\epsilon_1p_m\alpha_kN_{rx}}{\epsilon_1(\sigma^2_{t,k} + \frac{p_m}{N_{rx}}\sum\limits^S_{j\neq k} \alpha_j\phi_{s,kj}^2+ 1 )+ (1-\epsilon_1)(\sigma^2_{t,k} +p_m\alpha_k + 1)} \Bigg),
%\end{equation}
%
%
%
where $\sigma^2_{t,k}=p_s\sigma^2_{s,k} +p_s\sum^S_{j\neq k}\sigma^2_{c,kj}$. Furthermore, let us assume that, for the first phase, $p_m=E_m/N_{rx}$ and $p_s=E_s/M_{rx}$, where $E_m$ and $E_s$ are kept fixed regardless of $M_{rx}$ and $N_{rx}$. Then, as $M_{rx}\rightarrow \infty$ and $N_{rx}\rightarrow \infty$ with the same speed, the SEs at the MC BS and the $k$-th SC BS, respectively, approach 
\setcounter{equation}{27}
\begin{align}\label{eqinfrecmc1}
&\tilde{R}^{(1)}_{m,k} = \tau_d\log_2(1 + \rho_1E_s\beta_k),
\end{align}
\begin{equation}\label{eqinfrecsc1}
\tilde{R}^{(1)}_{s,k} = \tau_d\log_2(1 + \epsilon_1E_m\alpha_k).
\end{equation}

\textit{Remark 2}: For the second phase, as $K\rightarrow \infty$, the SE achieved at the MC BS and the $k$-th SC BS, converge to \eqref{eqrateconvergemc2} and \eqref{eqrateconvergesc2}, respectively. 
\begin{equation}
\label{eqrateconvergemc2}
\tilde{R}^{(2)}_{m,k} = \tau_d\log_2\Bigg(1 +   \dfrac{\rho_2p_s\alpha_kN_{tx}}{\rho_2\gamma_{m,k}^{(2)}+ (1-\rho_2)(p_sN_{tx}\alpha_k + \gamma_{m,k}^{(2)})} \Bigg), 
\end{equation}
%%%
\begin{equation}
\label{eqrateconvergesc2}
\tilde{R}^{(2)}_{s,k} = \tau_d\log_2\Bigg(1 +   \dfrac{\epsilon_2p_m\beta^2_kM_{tx}S}{\epsilon_2\gamma_{s,k}^{(2)}+ (1-\epsilon_2)(p_m\beta^2_kM_{tx}S + \gamma_{s,k}^{(2)})} \Bigg),
\end{equation}
where 
%\newline
$\gamma_{m,k}^{(2)}=p_m\zeta^2_{m,k}S + (p_s/N_{tx})\alpha_k\sum^S_{j\neq k} \psi_{m,kj}^2+ 1$, and 
%\newline
$\gamma_{s,k}^{(2)}=\sum^S_{j=1}\beta_j(p_s\zeta^2_{s,k} + p_s\sum^S_{j\neq k}\zeta_{c,kj}^2 + 1)+ (p_mS\beta_k/M_{tx})\sum^S_{j\neq k}\beta_j \psi_{s,kj}^2$.
Moreover, let us set the transmit powers  as $p_s=E_s/N_{tx}$ and $p_m=E_m/M_{tx}$ and then allow $N_{tx}\rightarrow \infty$ and $M_{tx}\rightarrow \infty$, irrespective of the value of $E_s$ and $E_m$. The SE at the MC BS and $k$-th SC BS approach saturated values given, respectively, by 
\begin{equation}\label{eqinfrecmc2}
\tilde{R}^{(2)}_{m,k} = \tau_d\log_2\Bigg(1 + \dfrac{\rho_2E_s\alpha_k}{1 + (1-\rho_2)E_s\alpha_k }\Bigg),
\end{equation} 
\begin{equation}\label{eqinfrecsc2}
\tilde{R}^{(2)}_{s,k} =\tau_d\log_2\Bigg(1 + \frac{\epsilon_2E_m\beta^2_k}{\frac{1}{S}\sum\limits_{j=1}^S\beta_j + (1-\epsilon_2)E_m\beta_k^2}\Bigg).
\end{equation} 

Remarks 1 and 2 show that as the line-of-sight components increase in strength, the SE approaches a constant  value which is independent of the Rician $K$-factor --- revealing that there is SE limit in strong LoS conditions. Furthermore, the results indicate that, for a fixed channel estimation accuracy and ADC resolution, the available transmission power can be scaled down according to the massive receive antennas (in the first phase) and the massive transmit antennas (in the second phase) and still achieve a non-zero rate. It is easy to see that as the ADC resolution $b\rightarrow \infty$, $\kappa\rightarrow 0$ and $\rho_1$, $\epsilon_1$, $\rho_2$, and $\epsilon_2$ approach 1. Thus, \eqref{eqinfrecmc1},  \eqref{eqinfrecsc1},   \eqref{eqinfrecmc2}, and  \eqref{eqinfrecsc2} reduce to Corollaries 1 and  2 of \cite{Anokye18}. 
Although as $b\rightarrow \infty$, the SE increases, the power consumption of the ADCs increase exponentially and therefore, places a huge burden on the EE of the system. In the next section, we will present an EE analysis by utilizing a practical power consumption model and then demonstrate the validity of our analyses. 

\section{Numerical Results}\label{secresults}
This section presents the results of the paper. Analytic results are validated through Monte Carlo simulations over $10^5$ channel realizations. The SNR is defined as $\text{SNR}\overset\Delta{=}p_m$ and $p_m=p_s$, i.e., we assume that the noise has unit power. Unless otherwise stated, the following parameters are used throughout: For the first phase, the large-scale fading coefficients are set as $\beta_k=\beta=0.2, \hspace{0.1cm} \forall k$ and $\alpha_k=\alpha=0.2, \hspace{0.1cm} \forall k$. The SI coefficient at the MC BS and the SC BSs are, respectively, given by $\sigma^2_m=0.3$ and $\sigma_{s,k}^2=\sigma^2_{s}=0.3, \hspace{0.1cm}\forall k$. The SC-to-SC interference strength from the $j$-th SC BS to the $k$-th SC BS is set as $\sigma_{c,kj}^2=\sigma_c^2=0.2, \hspace{0.1cm}\forall k,j$. In the second phase, the SI at the MC BS and the SC BSs are $\zeta_{m,k}^2=\zeta_m^2=0.3, \hspace{0.1cm}\forall k$ and $\zeta_{s,k}^2=\zeta_{s}^2=0.3, \forall k$, respectively, and the SC-to-SC interference at the SC BSs is $\zeta_{c,kj}^2=\zeta_{c}^2=0.2, \hspace{0.1cm} \forall  k,j$. Also, we have $S=6$, $p_\tau=p_m=p_s=10\textrm{dB}$, $T=200$ and $\tau_p=2S$. 
For simplicity, we set the Rician $K$-factor as $K_{m,k}=K_m, \hspace{0.1cm}\forall k$, $K_{s,k}=K_s, \hspace{0.1cm} \forall k$, and $K_m=K_s=K$ with $K=0\text{dB}$. To obtain $\rho_1$, $\epsilon_1$, $\rho_2$, and $\epsilon_2$, see Table \ref{tablebits}. 
%The simulation parameters are summarized in Table \ref{table:parameters}.
%

\subsection{Spectral Efficiency}
\begin{figure*}[t!]%\vspace{-0.4cm}
\centering
\includegraphics[width=14.0cm,height=10.0cm,keepaspectratio]{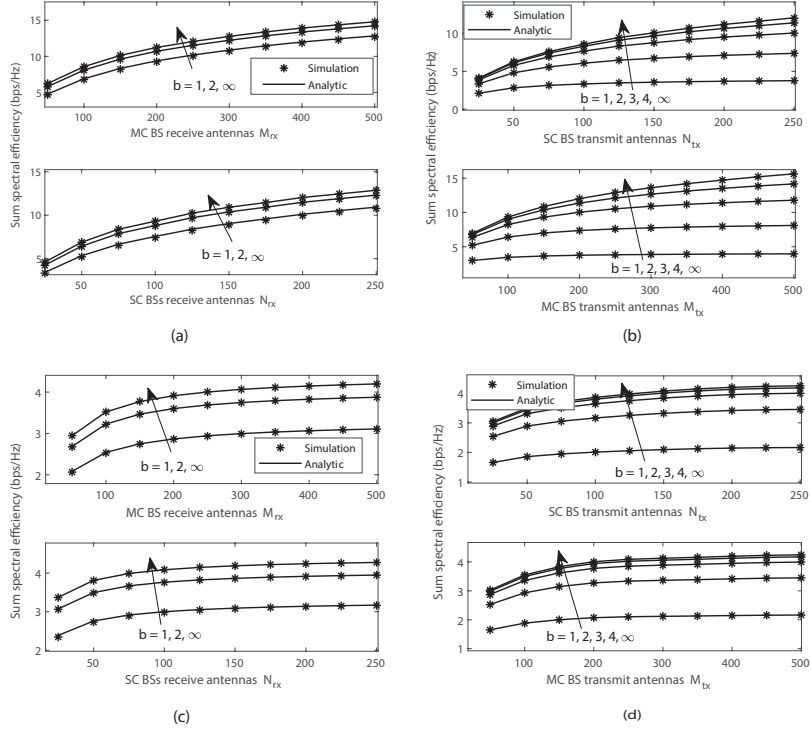}
\caption{ (a) First phase: Sum SE vs number of receive antennas ($\beta=\alpha=0.2$, $\sigma^2_m=\sigma_s^2=0.3$, $\sigma^2_c=0.2$; $K=0\text{dB}$, $p_m=p_s=10\text{dB}$, $S=6$, $\tau_p=2S$). (b) Second phase: Sum SE vs number of transmit antennas ($\zeta^2_m=\zeta_s^2=0.3$, $\zeta^2_c=0.2$). (c) First phase: Sum SE vs receive antennas ($p_m=\frac{E_m}{N_{rx}}$, $p_s=\frac{E_s}{M_{rx}}$, $E_m=E_s=10\text{dB}$) (d) Second phase: Sum SE vs transmit antennas ($p_m=\frac{E_m}{M_{tx}}$, $p_s=\frac{E_s}{N_{tx}}$).}
\label{figvsrecant}%\vspace{-0.5cm}
\end{figure*}
This subsection shows the sum SE  of the first and second phases. The  ``Analytic'' results are obtained by using \eqref{eqrateanalmc1} - \eqref{eqrateanalsc2} while ``Simulation'' represents the Monte Carlo simulations by utilizing \eqref{eqratemc1} and \eqref{eqratemc2}. 
Fig. \ref{figvsrecant}(a) illustrates the sum SE vs the number of receive antennas, for the first phase. The upper and lower subplots show the UL sum SE at the MC BS (using \eqref{eqrateanalmc1})  and the DL sum SE at the SC BSs (with \eqref{eqrateanalsc1}), respectively. The results are then aggregated over all $S$ SC BSs. As observed, the analytic results are highly accurate when compared to the Monte Carlo simulations. The sum SE increases without bound as the number of receive antennas increases. The sum SE deteriorates as the ADCs' resolution reduces. 
However, the performance loss due to the QN is very low in comparison to the infinite-resolution ADC. For example, at the MC BS and with $M_{rx}=200$, the 2-bit resolution case is able to achieve about 95\% of the sum SE with respect to the perfect ADC.   Again, the limited SE loss due to the QN can be compensated by simply increasing the  antennas. 
For instance, with infinite-resolution ADCs, the MC BS requires about 150 receive antennas to attain a sum SE of 10 bps/Hz and with a 2-bit ADC resolution, it needs about 200 receive antennas to achieve similar sum SE.

For the second phase, Fig. \ref{figvsrecant}(b) shows a plot of the sum SE against the number of transmit antennas. The first subplot represents the sum SE at the MC BS against the massive transmit antennas employed by the SC BSs (i.e., according to \eqref{eqrateanalmc2}). The second subplot shows the sum SE at the SC BSs versus the massive transmit antennas utilized at the MC BS (using \eqref{eqrateanalsc2}). The approximations are tight relative to the Monte Carlo simulations. Different from the first phase, the sum SE in the case of the low-resolution ADCs, saturates rapidly as the transmit antennas increase in the second phase. According to Lemma 2, the desired signal and the QN have the same order. Therefore, at low-resolution ADCs, the QN becomes more pronounced and a performance ceiling is introduced.

Next, we investigate the potential to scale down the available transmit power while achieving a given quality-of-service (QoS). We set  $p_m=\frac{E_m}{N_{rx}}$, $p_s=\frac{E_s}{M_{rx}}$, and $E_m=E_s=10\text{dB}$. Fig. \ref{figvsrecant}(c) illustrates a plot of the sum SE versus the number of receive antennas, for the first phase. The sum SE improves as the number of receive antennas increases until saturation. 
This demonstrates that, for any fixed channel estimation accuracy, we can scale down the available transmit power at the BSs 
with the massive receive antennas and still achieve a nonvanishing SE. This is consistent with the Remark 1 (i.e., equation \eqref{eqinfrecmc1}). Again, the sum SE improves as the ADCs resolution increases. Fig. \ref{figvsrecant}(d) also demonstrates the power scaling for the second phase, for different quantization bits $b$. Here, $p_m=E_m/M_{tx}$, $p_s=E_m/N_{tx}$, and $E_m=E_s=10\text{dB}$. Similarly, we can cut down the transmit power of the MC BS proportional to the massive transmit antennas $M_{tx}$ and the SC BSs transmit power according to the massive transmit antennas $N_{tx}$ and still attain a given level of QoS. However, at low-resolution ADCs, the sum SE saturates more rapidly relative to the infinite-resolution ADCs' case.   
\begin{figure}[!htbp]%\vspace{-0.4cm}
\centering
\includegraphics[width=7.0cm,height=7.0cm,keepaspectratio]{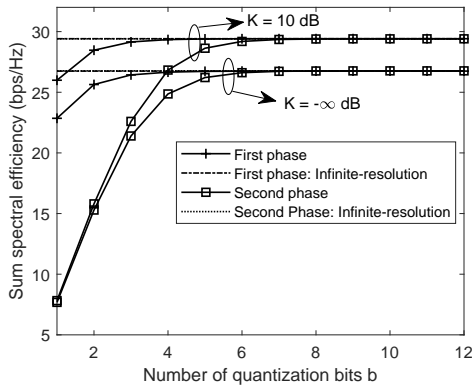}
\caption{Sum SE vs  quantization bits $b$ ($\beta=\alpha=0.2$, $\sigma^2_m=\sigma_s^2=0.3$, $\sigma^2_c=0.2$, $p_m=p_s=10\text{dB}$).}
\label{figvsbits}%\vspace{-0.5cm}
\end{figure}

To further understand the influence of the ADCs' resolution, a plot of the sum SE versus the number of quantization bits $b$ is shown in Fig. \ref{figvsbits}. For the purpose of comparison, we include the results for the Rayleigh fading scenario (corresponding to $K=-\infty \text{dB}$) and the infinite-resolution ADC case. For the first phase, the receive antennas are set as $M_{rx}=500$ and $N_{rx}=250$. In the case of the second phase, we use $M_{tx}=500$ and $N_{tx}=250$, due to the switching of the  antenna roles.
For both phases, we add the UL/DL SE and then aggregate over all the $S$ SCs, i.e., we have $\sum^S_{j=1}(R^{(1)}_{m,j}+R^{(1)}_{s,j})$ and $\sum^S_{j=1}(R^{(2)}_{m,j}+R^{(2)}_{s,j})$, for the first and second phases, respectively. Due to the tight approximation, we drop the Monte Carlo simulations. From the figure, the sum SE is marginally enhanced as the Rician $K$-factor increases. Furthermore, the sum SE improves as the quantization bits $b$ increases until it converges to a constant value (equal to the perfect ADC), for both the first and second phases. However, in the low-resolution ADCs' regime, the sum SE of the first phase outperforms the second phase.
This is because with the low-resolution ADCs, the QN  becomes more significant in the second phase and the SE deteriorates. The sum SE also generally improves with the growth in the Rician $K$-factor. It is also instructive to note that, for the second phase, the SE performance gap between the case of $K=10$ dB and  $-\infty$ dB is indistinguishable at the low-resolution ADCs' regime, i.e., from $b=1$ to 2 bits. 
From the Lemma 2, the QN and the desired signal have the same order and therefore, at low-resolution ADC, the QN becomes more pronounced and dominates the SE performance. However, as the quantization bits increases, the 
QN becomes less pronounced and the SE performance gap at the different Rician $K$-factor values, i.e., $K=-\infty$dB and $K=10$dB, widens from $b=3$ to $b=6$. 
At $K=10\text{dB}$, the first phase requires about 4bits ADC resolution per receive antenna to achieve equal performance as the infinite-resolution while the second phase requires approximately 7bits ADC resolution. It is worth noting that far fewer receive antennas are used in the second phase. 
\begin{figure}[t!]%\vspace{-0.5cm}
\centering
\includegraphics[width=7.0cm,height=7.0cm,keepaspectratio]{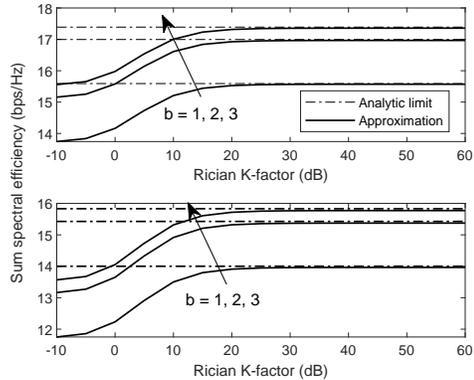}%\vspace{-0.4cm}
\caption{Sum SE vs Rician $K$-factor ($\beta=\alpha=0.2$, $\sigma^2_m=\sigma_s^2=0.3$, $\sigma^2_c=0.2$; $M_{rx}=700$, $N_{rx}=350$, $p_m=p_s=10\text{dB}$).}
\label{figvsK1st}%\vspace{-0.5cm}
\end{figure}

To elaborate the implications of the Rician $K$-factor, we plot a graph of the sum SE versus Rician $K$-factor in Fig. \ref{figvsK1st}, where $M_{rx}=700$ and $N_{rx}=350$. The upper subplot shows the SE at the MC BS and the lower subplot indicates the sum SE at the SC BSs, for the first phase. Also, the ``Approximation'' is obtained by using the  Lemma 1.  From the curves, it is observed that the sum SE increases with the Rician $K$-factor at both the MC BS and the SC BSs until it converges to a saturation value indicated by the ``Analytic limit'' plot (which is derived in Remark 1,  \eqref{eqrateconvergemc1} and \eqref{eqrateconvergesc1}).  For brevity, we have omitted the results for the second phase.

\begin{figure*}
\centering
\begin{subfigure}{.5\textwidth}
  \centering
  \includegraphics[width=0.90\linewidth]{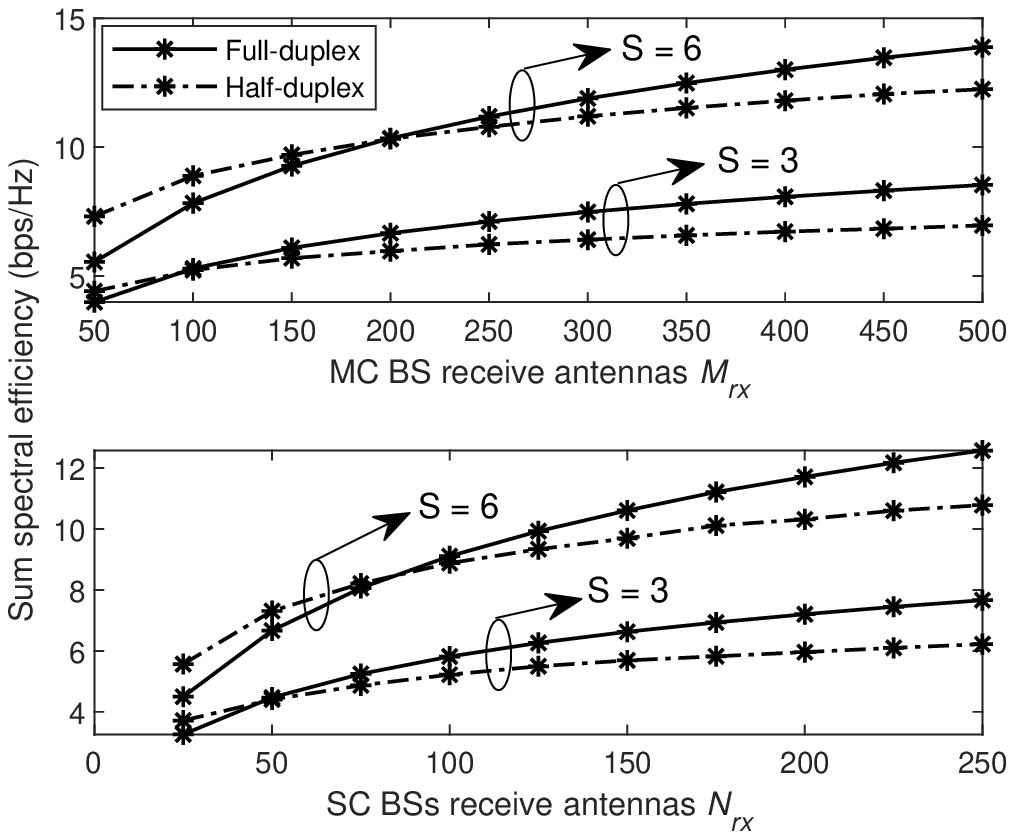}%\vspace{-0.4cm}
  \caption{}
\end{subfigure}%
\begin{subfigure}{.5\textwidth}
  \centering
  \includegraphics[width=0.90\linewidth]{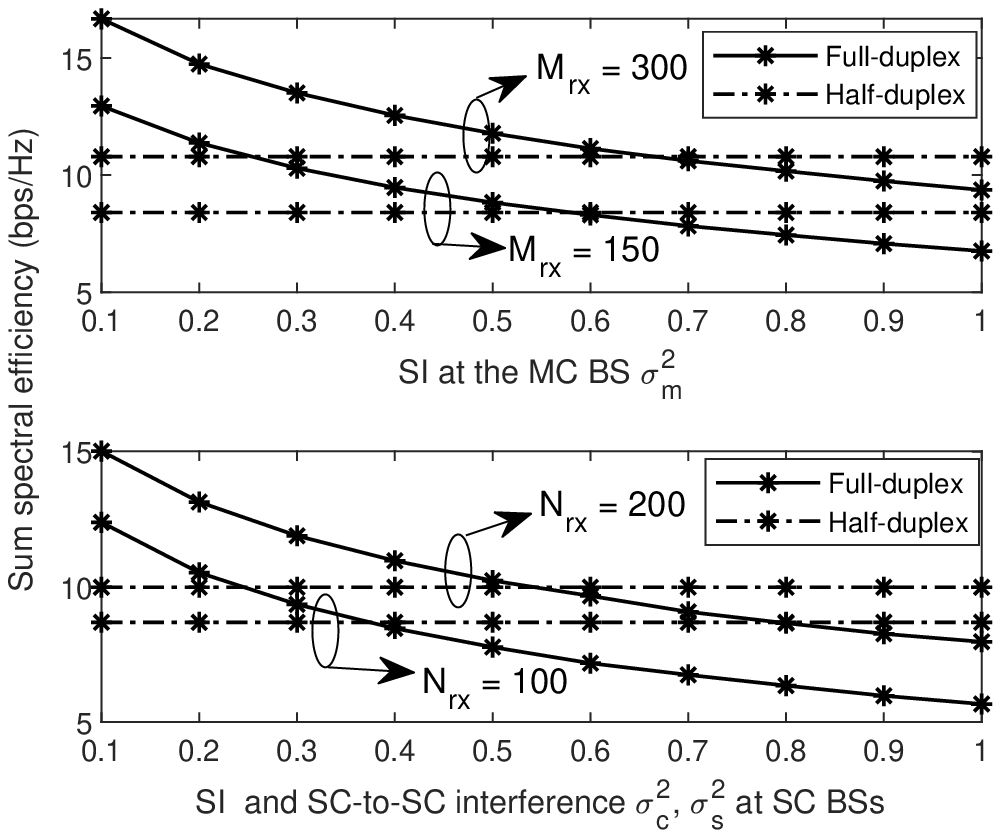}%\vspace{-0.4cm}
  \caption{}
\end{subfigure}%\vspace{-0.6cm}
\caption{(a) Sum SE vs number of receive antennas ($\beta=\alpha=0.2$, $\sigma^2_m=\sigma_s^2=0.5$, $\sigma_c^2=0.3$,  $K=10\text{dB}$, $p_m=p_s=10\text{dB}$, $b=3$, and $\tau_p=2S$). (b) Sum SE vs SI (and SC-to-SC interference).}
\label{figvsantscs}%\vspace{-0.6cm}
\end{figure*}

Fig. \ref{figvsantscs} demonstrates the implications of the SI and SC-to-SC interference on the sum SE, for the first phase.  For comparison, we have included the HD case. The  HD  SE is obtained by  setting the SI and SC-to-SC interference coefficients to zero, doubling the transmit powers and including a pre-log factor of $1/2$ in the rate equation. Fig.  \ref{figvsantscs}(a) shows the sum SE versus the receive antennas, where we set $K=10$dB, $b=3$, $\sigma_m^2=\sigma_s^2=0.5$, and $\sigma^2_c=0.3$. As expected, the sum SE generally enhances as the number of SC BSs increases. The HD system outperforms the FD in the low receive antenna regime. However, as the receive antennas increase, the FD begins to show superior performance. This is because the massive antennas asymptotically null the influence of the SI and SC-to-SC interference. Furthermore, it is observed that more receive antennas are needed for the FD  to outperform its HD counterpart as the SC BSs increase. 
This happens because the number of SC BSs enhances the potency of the SI and SC-to-SC interference (and increases the strength of the QN), as proved in the Lemmas 1 and 2. In Fig. \ref{figvsantscs} (b), we plot a graph of the sum SE against the SI and SC-to-SC interference coefficients. For the MC BS, we vary the SI coefficient $\sigma^2_m$ whereas in the case of the SC BSs, we jointly vary the SI $\sigma^2_s$ and SC-to-SC interference $\sigma_c^2$ coefficients. As observed, the sum SE of the FD decays rapidly as the SI  and SC-to-SC interference increase. The sum SE of the HD case remains unchanged since it is independent of the SI and SC-to-SC interference. The tolerance of the FD system to the SI and SC-to-SC interference impacts is enhanced by the higher number of receive antennas. For example, at the MC BS, with 150 receive antennas, the FD system is able to withstand SI strength of about 0.56 and increases to 0.68 when 300 receive antennas are employed.  
%Similar analysis can be extended to the second phase but it is omitted here due to space constraints.
%
%%
%\begin{figure*}[!htbp]%\vspace{-0.4cm}
%\centering
%\includegraphics[width=12.0cm,height=6cm,keepaspectratio]{/fdhd_compare.eps}%\vspace{-0.4cm}
%\caption{(a) Sum SE vs number of receive antennas ($\beta=\alpha=0.2$, $\sigma^2_m=\sigma_s^2=0.5$, $\sigma_c^2=0.3$,  $K=10\text{dB}$, $p_m=p_s=10\text{dB}$, $b=3$, and $\tau_p=2S$). (b) Sum SE vs SI (and SC-to-SC interference).}
%\label{fig:vs_ant_scs}%\vspace{-0.4cm}
%\end{figure*}
%%%
\subsection{Energy Efficiency}
From the previous sections, the unquantized system outperforms the low-resolution ADC quantized counterpart in all the receive antenna regimes. Specifically, a closer look at Fig. \ref{figvsrecant}(a) shows that we can compensate for the rate loss due to the QN simply by increasing  receive antennas.  However, an increase in the receive antennas increases the hardware cost and the power consumption since more ADCs will be subsequently needed.   
Also in Fig. \ref{figvsrecant}(b), it is observed that, for the second phase, in the low-resolution ADC regime, the sum SE saturates rapidly with the massive transmit antennas. In this case, therefore, high-resolution ADCs will be required. However, increasing the quantization bits further exacerbates the power consumption of the receive radio frequency (RF) chains during the second phase, which implies that a fundamental trade-off  exists between the SE and the power consumption
for the practical deployment of the proposed FD massive MIMO-enabled backhaul. To address such a trade-off issue, in this section, we investigate the performance of the proposed scheme in terms of EE at the MC BS and the SC BS receivers. Note that the transmit powers of the BSs are ignored. This is a common assumption  in the low-resolution ADC and mixed ADC analysis  \cite{ Mendez16}.

The EE is defined as \cite{Zhang2017}
\begin{equation}\label{eqee}
\Sigma^{(i)}_{EE} = \frac{B_wR^{(i)}_{SE}}{P^{(i)}_C} \hspace{0.2cm} \text{bit/Joule},
\end{equation}
where $B_w$ and $P_C^{(i)}=P^{(i)}_{C,m} + \sum^{S}_{k=1}P^{(i)}_{C,sk}$ denote the bandwidth and  the total power consumption, respectively. Also, $P^{(i)}_{C,m}$ and $P^{(i)}_{C,sk}$ indicate the power consumption at the MC BS and the $k$-th SC BS, respectively. $R^{(i)}_{SE}= \sum^S_{k=1}(R^{(i)}_{s,k} + R^{(i)}_{m,k})$  describes the total UL and DL sum SE; $i\in [1,2]$ denotes the first phase or second phase. 
To model the power consumption, we consider a generic power consumption model  for evaluating the low-resolution ADCs architecture \cite{Mendez16}. Thus, the total power consumption for the first phase is modeled as
\begin{align*}
&P_{C,m}^{(1)}=  M_{rx}(P_{LNA} + P_{RFC} + 2P_{ADC} ) + P_{BB},\hspace{0.2cm}\textrm{and}\\& P_{C,sk}^{(1)}=N_{rx}(P_{LNA} + P_{RFC} + 2P_{ADC} ) + P_{BB},
\end{align*}
%%
%\begin{align}\label{eq:pwr_consume}
%P_{C,sk}^{(1)}=N_{rx}(P_{LNA} + P_{RFC} + 2P_{ADC} ) + P_{BB}\hspace{0.1cm} ,
%\end{align}
%%
where  $P_{LNA}$, $P_{ADC}$, and $P_{BB}$ denote the power consumption in the low noise amplifier (LNA), ADC, and baseband (BB) processor, respectively. Also, $P_{RFC}=(P_{M} + P_{LO} + P_{LPF} + P_{Bamp})$ denote the total power consumed by the RF chain, where $P_M$, $P_{LO}$, $P_{LPF}$, and $P_{Bamp}$ are the power consumed by the mixer, local oscillator, low-pass filter (LPF), and baseband amplifier, respectively. An approximate total power consumed by the  RF chain, i.e., $P_{RFC}$, is 40mW \cite{Mendez16}. 
For the second phase, the number of receive antennas at the MC BS  are fixed such that $M_{rx}=S$ and a single receive antenna at the $k$-th SC BS. Therefore, the power consumption at the MC BS and the $k$-th SC BS are, respectively, written as 
\begin{align*}
&P_{C,m}^{(2)}=S(P_{LNA} + P_{RFC} + 2P_{ADC} ) + P_{BB}, \hspace{0.1cm}\textrm{and}\hspace{0.1cm}\\&P_{C,sk}^{(2)}=(P_{LNA} + P_{RFC} + 2P_{ADC} ) + P_{BB}\hspace{0.1cm}.
\end{align*}
Moreover, the power consumption of the ADC is related to the number of quantization bits $b$ by 
\begin{align}\label{eqpadc}
P_{ADC} = \text{FOM}_{W}\cdot f_s\cdot 2^b\hspace{0.1cm},
\end{align}
where $\text{FOM}_W$ and $f_s$ denotes the Walden's figure-of-merit  \cite{Walden97} and Nyquist sampling frequency, respectively. At 1 GHz bandwidth, $\text{FOM}_W$ takes approximate values of $5 \sim 65$ fJ/conversion-step \cite{Abbas17}.  Approximate power consumption of the other devices are $P_{LNA}=5.4$mW \cite{Shang12}, and $P_{BB}=200$mW \cite{Zhang2017}.  We set $\text{FOM}_W=15$ fJ/conversion-step, $B_w=1 \text{GHz}$, and $f_s=2B_w$. 
\begin{figure*}
\centering
\begin{subfigure}{.5\textwidth}
  \centering
  \includegraphics[width=0.90\linewidth]{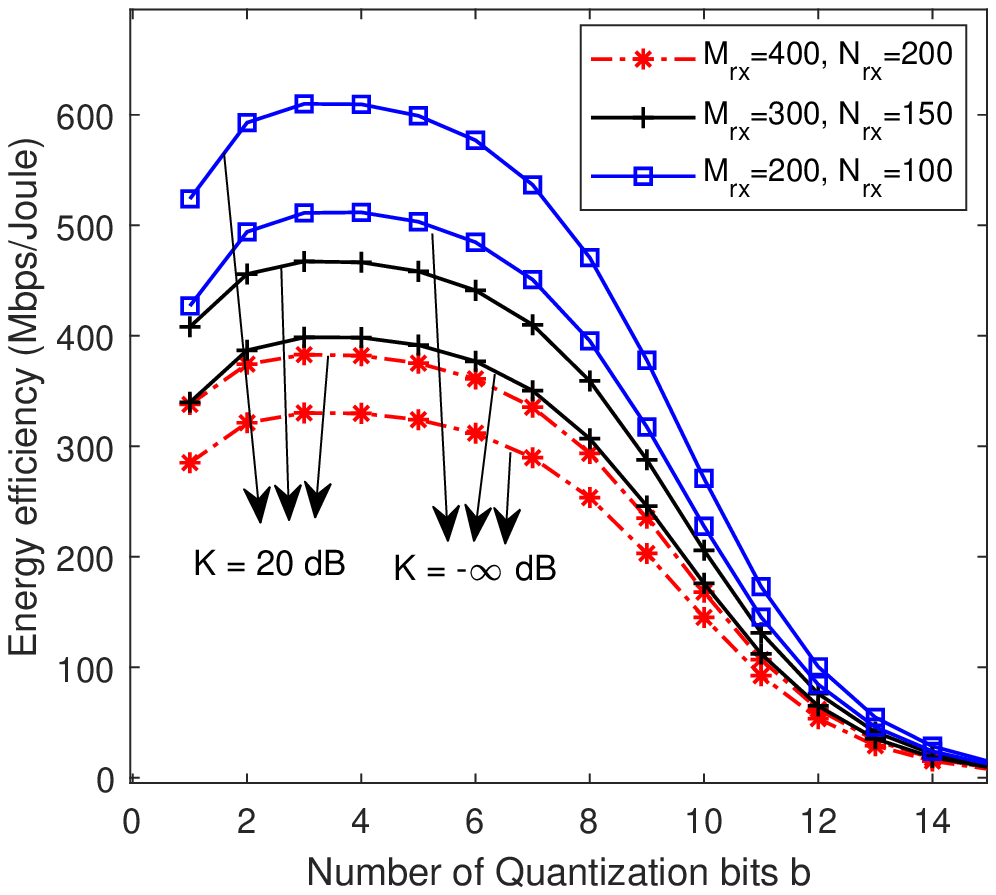}%\vspace{-0.4cm}
  \caption{}
  \label{figeebits1a}
\end{subfigure}%
\begin{subfigure}{.5\textwidth}
  \centering
  \includegraphics[width=0.90\linewidth]{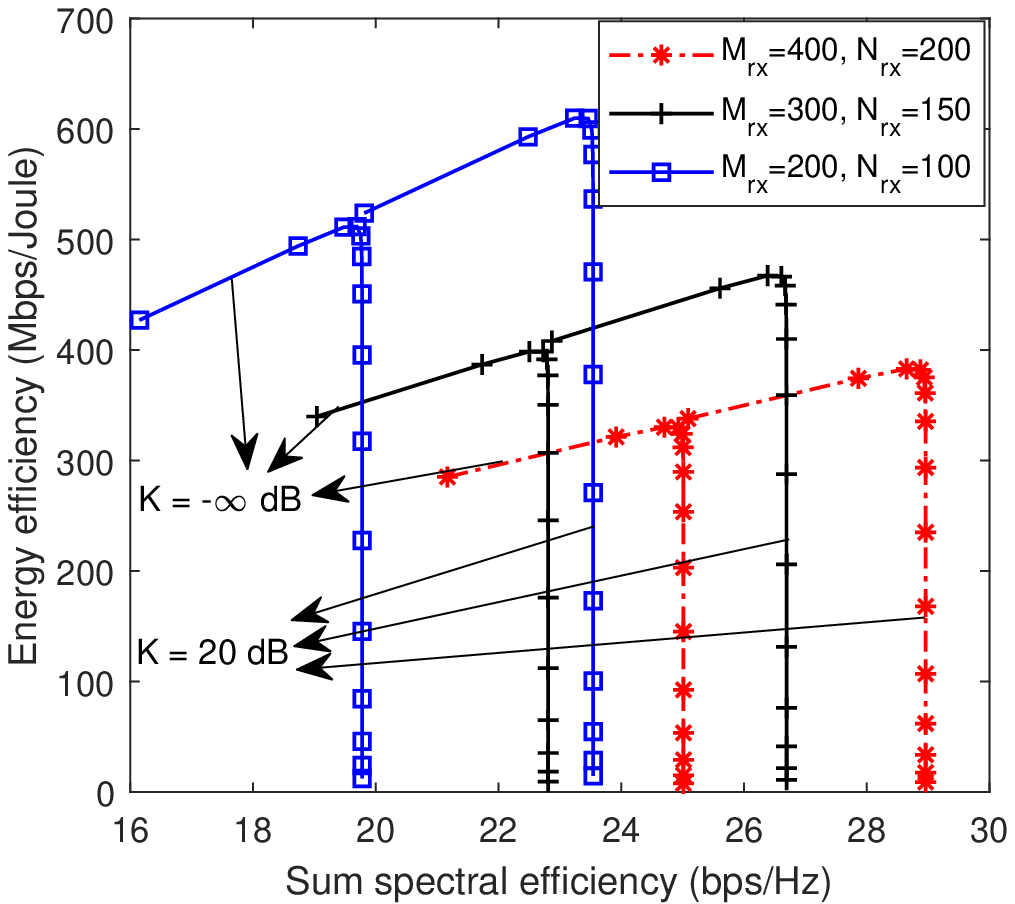}%\vspace{-0.4cm}
  \caption{}
  \label{figeese1a}
\end{subfigure}%\vspace{-0.6cm}
\caption{(a) First phase: EE (Mbs/Joule) vs number of quantization bits $b$ ($p_m=p_s=10\text{dB}$, and $S=6$). (b) Trade-off between EE (Mbs/Joule) and SE (bps/Hz)}
\label{figeebits}%\vspace{-0.6cm}
\end{figure*}
\begin{figure*}
\centering
\begin{subfigure}{.5\textwidth}
  \centering
  \includegraphics[width=0.90\linewidth]{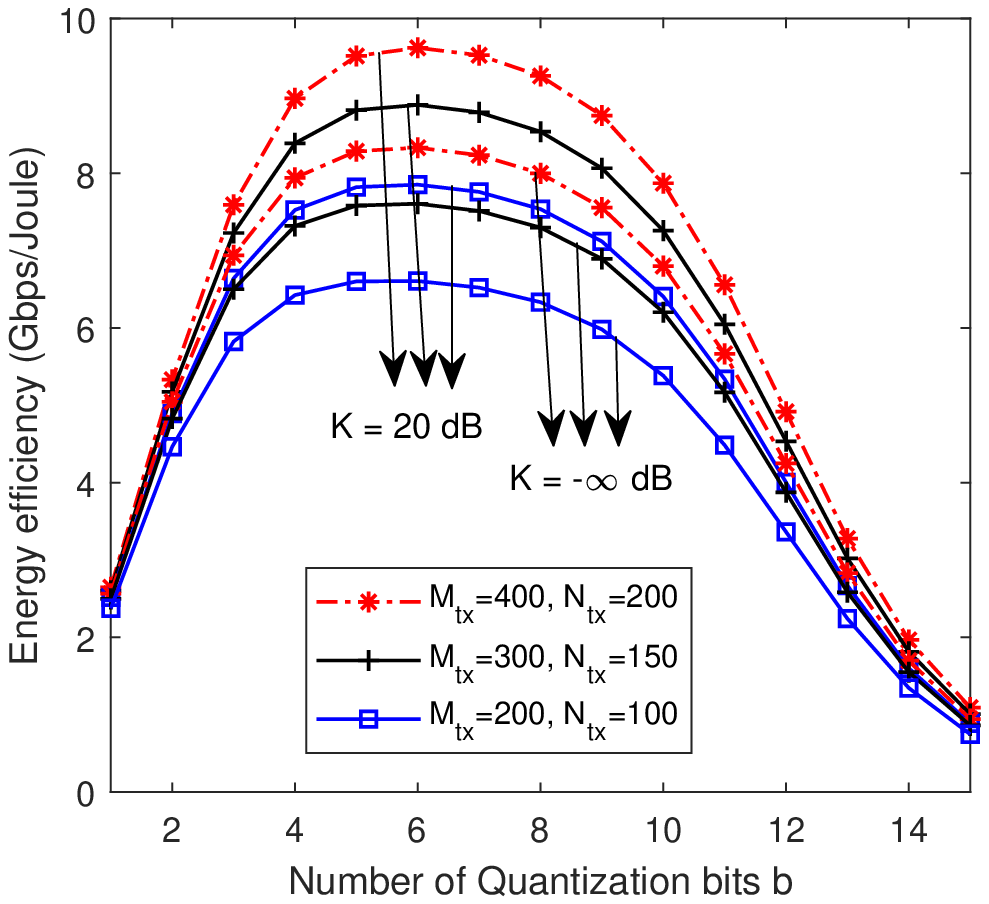}%\vspace{-0.4cm}
  \caption{}
  \label{figeebits1b}
\end{subfigure}%
\begin{subfigure}{.5\textwidth}
  \centering
  \includegraphics[width=0.90\linewidth]{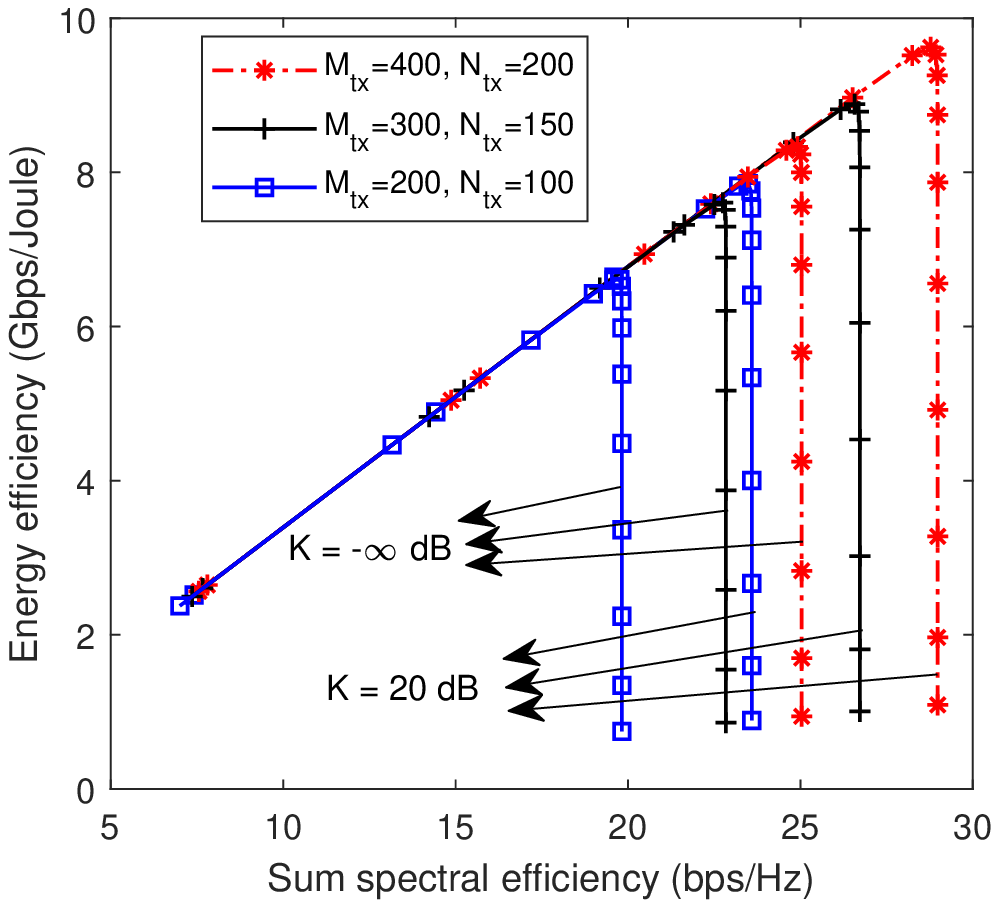}%\vspace{-0.4cm}
  \caption{}
  \label{figeese1b}
\end{subfigure}%\vspace{-0.6cm}
\caption{(a) Second phase: EE (Gbps/Joule) vs number of quantization bits $b$ ($p_m=p_s=10\text{dB}$, and $S=6$). (b) Trade-off between EE (Gbps/Joule) and SE (bps/Hz)}
\label{figeebits2nd}%\vspace{-0.6cm}
\end{figure*}

Using \eqref{eqee},  Fig. \ref{figeebits}(a) plots the EE (in Mbps/Joule) versus the  quantization bits $b$, for the first phase. The EE initially improves as the bits increases from $b=1$ to $b=3$  until the EE deteriorates towards 0 for $b>3$. This is because the sum SE improves sub-linearly as $b$ increases whereas the  ADC power consumption increases exponentially with $b$. Therefore, with a high-resolution ADC,  the power consumption dominates and the EE decreases. A simple line search algorithm such as bisection, can be used to obtain the optimal  bits which maximizes the EE. Also, the EE generally decreases with the increasing number of receive antennas since each requires a pair of ADCs. Therefore, the power consumption increases linearly with the number of receive antennas. 
It is observed that the EE is also enhanced with the growth in the Rician $K$-factor as the sum SE increases with the Rician $K$-factor (and the power consumption of the receivers is independent of the Rician $K$-factor).  
Fig. \ref{figeebits}(b) shows the EE/SE trade-off for different Rician $K$-factors, receive antennas (i.e., $M_{rx}$ and $N_{rx}$), and quantization bits $b$, varied from $b=1$ to 15. For each $b$, we compute the EE and the sum SE $R^{(1)}_{SE}$ (in bps/Hz). The superior sum SE values are shown in the rightmost points whereas the best EE values are obtained at the uppermost points. Thus, the best EE/SE trade-offs are achieved at  the right-uppermost points.  As illustrated in Fig. \ref{figeebits}(b), both the EE and the SE increase from $b=1$ to 3. 
However, a further increase in the ADC resolution reveals that  EE reduces rapidly while the SE remains stagnant. There is also an advancement in the sum SE and a deterioration in the EE with the increase in the receive antennas. Furthermore, as the Rician $K$-factor  increases, the envelope of the entire EE/SE region grows. Thus, the flexibility of the system as a function of the ADCs' resolution is enhanced.

\begin{figure*}
\centering
\begin{subfigure}{.5\textwidth}
  \centering
  \includegraphics[width=0.9\linewidth]{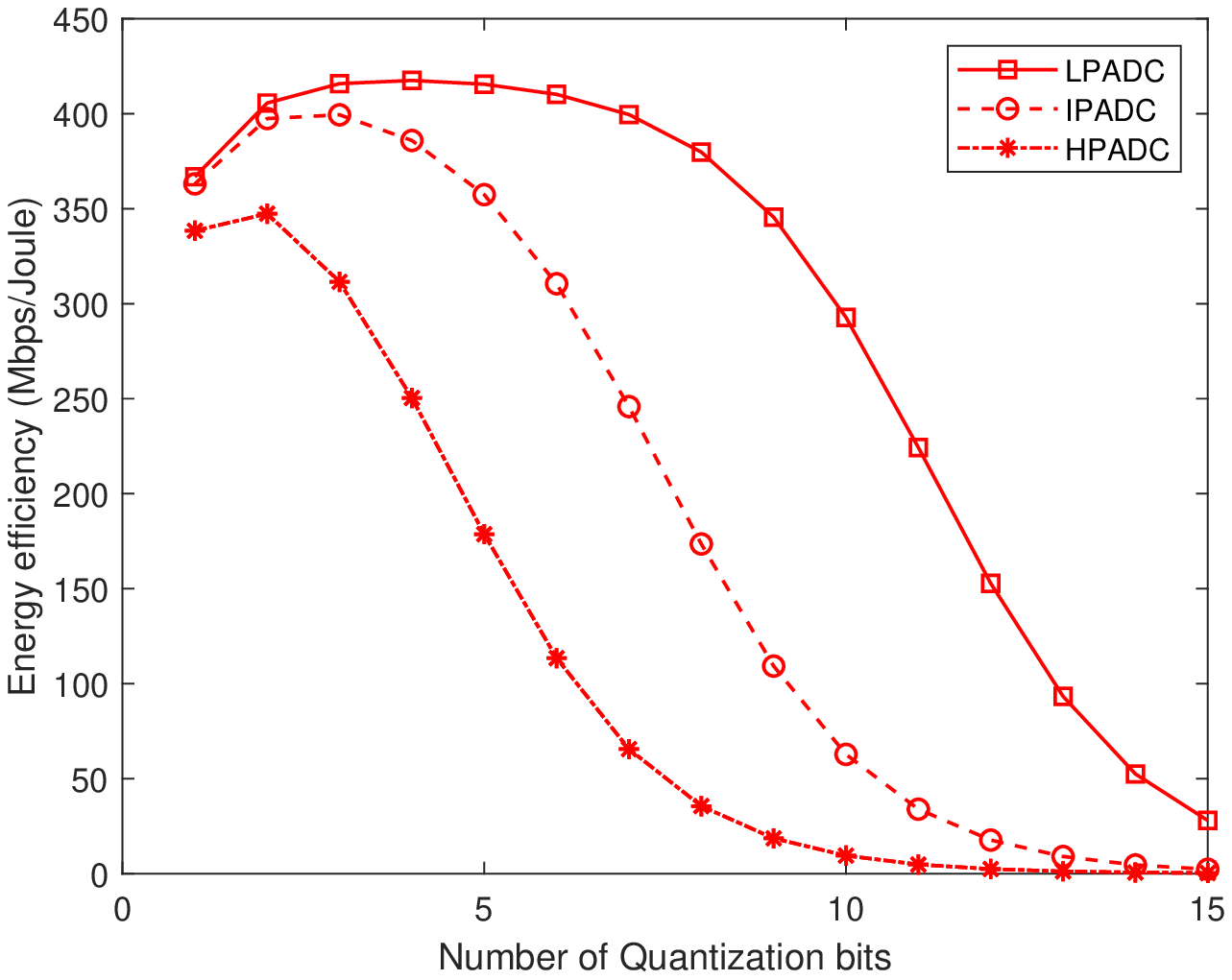}%\vspace{-0.4cm}
  \caption{}
  \label{figsub1}
\end{subfigure}%
\begin{subfigure}{.5\textwidth}
  \centering
  \includegraphics[width=0.9\linewidth]{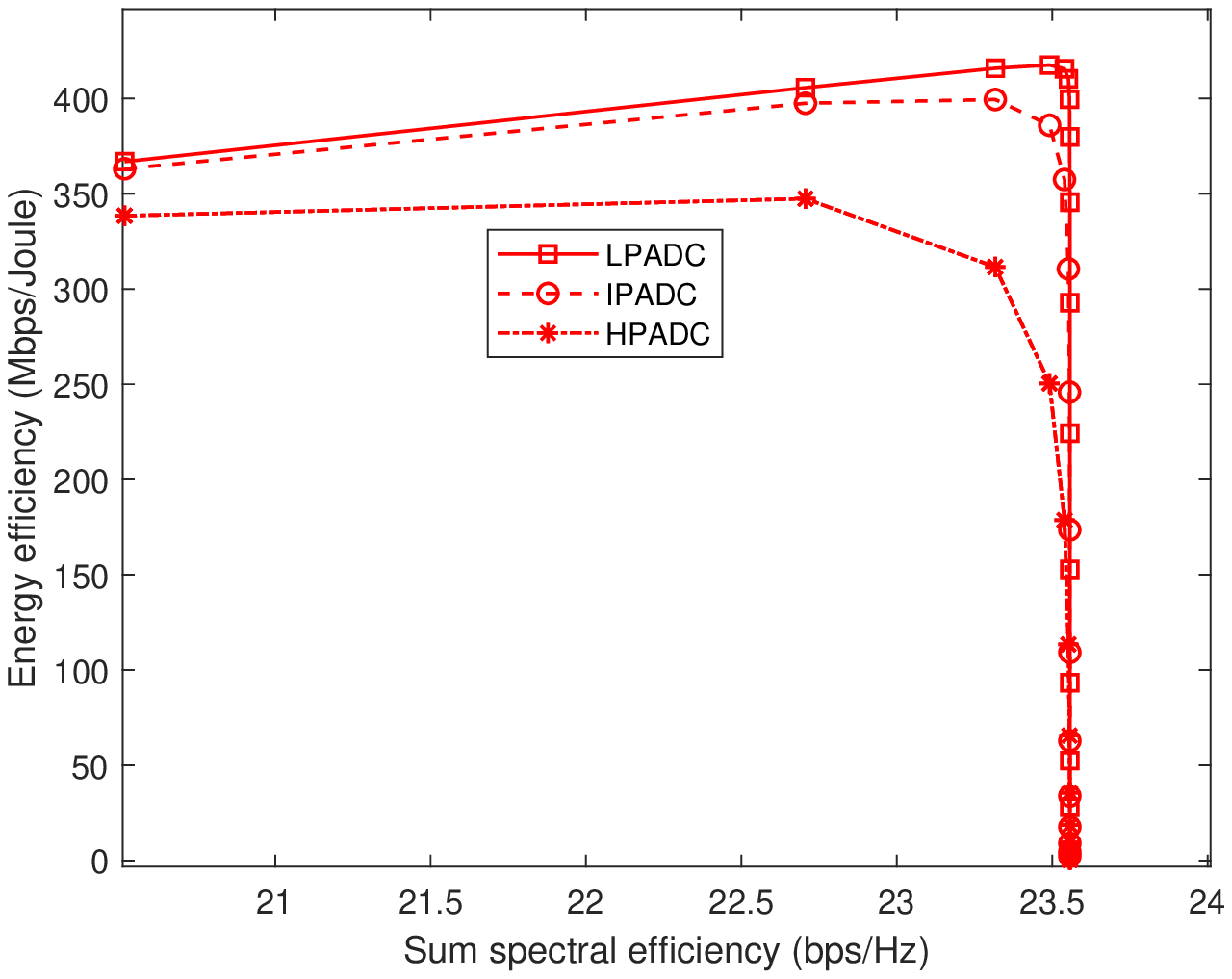}%\vspace{-0.4cm}
  \caption{}
  \label{figsub2}
\end{subfigure}%\vspace{-0.4cm}
\caption{First phase: (a) EE (Mbps/J) vs quantization bits for different $\text{FOM}_W$ ($S=6, p_m=p_s=10\text{dB}, M_{rx}=300, N_{rx}=200, )$ and $K=20\text{dB}$ (b)Trade-off between EE and sum SE }
\label{figeebitsfomw}%\vspace{-0.6cm}
\end{figure*}

For the second phase, Fig. \ref{figeebits2nd} provides insights into the EE/SE, for different massive transmit antennas (i.e., $M_{tx}$ and $N_{tx}$), Rician $K$-factor values, and quantization bits $b$. From Fig. \ref{figeebits2nd}(a), the EE  improves as $b$ increases from 1 to 6 until the EE begins to deteriorate for $b>6$. This is intuitive as the SE increases with the  quantization bits until it reaches a saturation point (see Fig. \ref{figvsbits}). Meanwhile, the power consumption increases exponentially with $b$. The EE appreciates with the Rician $K$-factor and the massive transmit antennas since the sum SE enhances with the Rician $K$-factor and the massive transmit antennas---the power consumption of the receiver is independent of the Rician $K$-factor and massive transmit antennas. 
Fig. \ref{figeebits2nd}(b) shows the EE/SE trade-off for the second phase.  We vary $b$ from 1 to 15 and for each $b$, the sum SE and the EE are calculated. The EE and the sum SE grow from $b=1$ to $b=6$. As $b$ grows beyond 6 bits, the EE monotonically decreases as the sum SE remains saturated. In contrast to the first phase, the envelope of the EE/SE region, for the different Rician $K$-factors and massive transmit antennas, are only distinguishable at high-resolution region.
Furthermore, we note that  the second phase EE is several orders superior to the first phase since far fewer receive antennas are used. Therefore, it is economically feasible to use high-resolution ADCs during the second phase.

From \eqref{eqpadc}, the ADCs' power consumption  grows linearly with the $\text{FOM}_W$ whose value has been rapidly changing over the past decade as observed in \cite{Abbas17}. In Fig. \ref{figeebitsfomw}, we plot the EE versus the quantization bits and the EE/SE trade-off curves for different $\text{FOM}_W$ values which correspond to low power (LP) ADC ($\text{FOM}_W=5\text{fJ/step/Hz}$), intermediate power (IP) ADC ($\text{FOM}_W=\text{65fJ/step/Hz}$), and high power (HP) ADC ($\text{FOM}_W=\text{494fJ/step/Hz}$)\cite{Abbas17}. Here, we consider only the first phase, for brevity. From Fig. \ref{figeebitsfomw}(a), it is observed that the optimal bits changes with the FOM$_W$ values. For instance, with the LPADC, the EE is maximized with 4 bits whereas the optimal bits to maximize the EE is 2 bits for the HPADC. From Fig. \ref{figeebitsfomw}(b), the EE/SE  region grows with reduction in the FOM$_W$ values. Thus improving the system's versatility.

\section{Conclusion}\label{secconc}
We have analyzed the sum SE and EE of a HetNet with backhaul enabled by low-resolution ADC quantized wireless FD massive MIMO over Rician channels. 
We have characterized the joint impact of the Rician $K$-factor, QN, SI, and SC-to-SC interference. During the first phase, the sum SE generally increases with the massive receive antennas but decreases with the reduction in ADCs' resolution. 
%However,  the SE loss due to QN is compensated for by employing massive receive antennas. 
On the contrary, in the second phase, the sum SE saturates rapidly as the massive transmit antennas increase in the low-resolution ADCs' region. Furthermore, it has been demonstrated that the sum SE  enhances with the increase in the Rician $K$-factor. We show that with a large-scale antenna,   the available power can be scaled down according to the massive receive antennas (first phase) and massive transmit antennas (second phase) and still achieve a nonvanishing SE. 
%This saturated SE is influenced only by the large-scale fading and the ADC resolution. 
A study into the EE performance shows that the EE initially increases with the ADCs' resolution  but deteriorates quickly as the quantization bits increase beyond a certain threshold.  The EE/SE trade-off region improves with the Rician $K$-factor.

%Numerically, it is shown that  the EE of the second phase is several orders superior to the first phase.  

% if have a single appendix:
%\appendix[Proof of the Zonklar Equations]
% or
%\appendix  % for no appendix heading
% do not use \section anymore after \appendix, only \section*
% is possibly needed

% use appendices with more than one appendix
% then use \section to start each appendix
% you must declare a \section before using any
% \subsection or using \label (\appendices by itself
% starts a section numbered zero.)
%

\appendices
\section{Proof of Lemma 1}
To prove \eqref{eqrateanalmc1}, we derive only the terms corresponding to the channel estimation error, SI, and QN, given by  $\rho_1^2p_s\sum^S_{j=1}\mathbb{E}[|\hat{\mathbf{h}}^H_k\mathbf{e}_j|^2]$,  $\rho_1^2p_m\mathbb{E}[|\hat{\mathbf{h}}^H_k\mathbf{Q}|^2]$ and $\mathbb{E}\big[\|\hat{\mathbf{h}}^H_k\mathbf{n}_q\|^2 ]$, respectively. The remaining terms can be obtained by using the results in \cite{Zhang14}. 
For the channel estimation error, 
\begin{align*}
\rho_1^2p_s\sum^S_{j=1}\mathbb{E}[|\hat{\mathbf{h}}^H_k\mathbf{e}_j|^2] = \rho_1^2p_s\sum^S_{j=1}\mathbb{E}[|\hat{\mathbf{h}}^H_k\mathbf{e}_j\mathbf{e}_j^H\hat{\mathbf{h}}_k|],
\end{align*}
where $\mathbb{E}[|\hat{\mathbf{h}}^H_k\mathbf{e}_j\mathbf{e}_j^H\hat{\mathbf{h}}_k|]= \tilde{\beta_j}\mathbb{E}[\|\hat{\mathbf{h}}_k\|^2]$. Also, by employing the law of large numbers and  \cite[Lemma 4]{Zhang14}, $\mathbb{E}[\|\hat{\mathbf{h}}_k\|^2]\rightarrow \frac{\beta_k}{K_{m,k}+1}(K_{m,k}+\eta_k)M_{rx}$ as $M_{rx}\rightarrow \infty$,  almost surely. Thus, 
\begin{align*}
\rho_1^2p_s\sum^S_{j=1}\mathbb{E}[|\hat{\mathbf{h}}^H_k\mathbf{e}_j|^2] = \frac{\beta_k}{K_{m,k}+1}(K_{m,k}+\eta_k)M_{rx}\rho_1^2p_s\sum^S_{j=1}\tilde{\beta_j}. 
\end{align*}

To derive the SI term, we have $\mathbb{E}[|\hat{\mathbf{h}}^H_k\mathbf{Q}|^2] =  \mathbb{E}[|\hat{\mathbf{h}}^H_k\mathbf{QQ}^H\hat{\mathbf{h}}_k|]=M_{tx}\sigma^2_m\mathbb{E}[\|\hat{\mathbf{h}}_k\|^2]$.
%%
%\begin{align*}
%&\mathbb{E}[|\hat{\mathbf{h}}^H_k\mathbf{Q}|^2] =  \mathbb{E}[|\hat{\mathbf{h}}^H_k\mathbf{QQ}^H\hat{\mathbf{h}}_k|]=M_{tx}\sigma^2_m\mathbb{E}[\|\hat{\mathbf{h}}_k\|^2],\\&
%\end{align*}
Therefore, the SI term is obtained as $\rho_1^2p_m\mathbb{E}[|\hat{\mathbf{h}}^H_k\mathbf{Q}|^2]=\rho_1^2p_mM_{tx}M_{rx}\sigma^2_m\frac{\beta_k}{K_{m,k}+1}(K_{m,k}+\eta_k)$.
%\begin{align*}
%\rho_1^2p_m\mathbb{E}[|\hat{\mathbf{h}}^H_k\mathbf{Q}|^2]=\rho_1^2p_mM_{tx}M_{rx}\sigma^2_m\frac{\beta_k}{K_{m,k}+1}(K_{m,k}+\eta_k).
%\end{align*}

For the QN, $\mathbb{E}\big[\|\hat{\mathbf{h}}^H_k\mathbf{n}_q\|^2 ]=\rho_1(1-\rho_1)\mathbb{E}[\hat{\mathbf{h}}_k^H\textrm{diag}(p_s\mathbf{HH}^H + p_m\mathbf{QQ}^H +\mathbf{I}_{M_{rx}})\hat{\mathbf{h}}_k].$
We can write 
\newline
$[\textrm{diag}(p_s\mathbf{HH}^H + p_m\mathbf{QQ}^H +\mathbf{I}_{M_{rx}})]_{nn}=p_s\sum^S_{j=1}|h_{nj}|^2 + p_m\sum^{M_{tx}}_{i=1}|q_{ni}|^2 + 1$. Therefore, 
$\mathbb{E}[\hat{\mathbf{h}}_k^H\times\textrm{diag}(p_s\mathbf{HH}^H+ p_m\mathbf{QQ}^H +\mathbf{I}_{M_{rx}})\hat{\mathbf{h}}_k] =\mathbb{E}\Big[ \sum^{M_{rx}}_{n=1}|\hat{h}_{nk}|^2+p_s\sum^{M_{rx}}_{n=1}|\hat{h}_{nk}|^2\sum^S_{j=1}|h_{nk}|^2 +\\ p_m\sum^{M_{rx}}_{n=1}|\hat{h}_{nk}|^2\sum^{M_{tx}}_{i=1}|q_{ni}|^2\Big]$.
%
%%
%%\vspace{-0.2cm}
%\begin{align}\label{eq:proof_qn}
%&\mathbb{E}[\hat{\mathbf{h}}_k^H\textrm{diag}(p_s\mathbf{HH}^H + p_m\mathbf{QQ}^H +\mathbf{I}_{M_{rx}})\hat{\mathbf{h}}_k] =\mathbb{E}\Big[ \sum^{M_{rx}}_{n=1}|\hat{h}_{nk}|^2+p_s\sum^{M_{rx}}_{n=1}|\hat{h}_{nk}|^2\sum^S_{j=1}|h_{nk}|^2 \nonumber\\&+ p_m\sum^{M_{rx}}_{n=1}|\hat{h}_{nk}|^2\sum^{M_{tx}}_{i=1}|q_{ni}|^2\Big].  
%\end{align}
%%
Using \eqref{eqcsimc}, the right hand side can be rewritten as
\begin{align*}
&\mathbb{E}\Big[\sum^{M_{rx}}_{n=1}|\hat{h}_{nk}|^2+p_s\sum^{M_{rx}}_{n=1}|\hat{h}_{nk}|^4 + p_s\sum^{M_{rx}}_{n=1}\sum^S_{j\neq k}|\hat{h}_{nk}|^2|\hat{h}_{nj}|^2 \\&+p_s\sum^{M_{rx}}_{n=1}\sum^S_{j= 1}|\hat{h}_{nk}|^2|\tilde{h}_{nj}|^2+ p_m\sum^{M_{rx}}_{n=1}\sum^{M_{tx}}_{i=1}|\hat{h}_{nk}|^2\times |q_{ni}|^2\Big]. 
\end{align*}
We can derive the following expectations:
\begin{align*}
&\mathbb{E}[|\hat{h}_{nk}|^2] = \frac{\beta_{nk}}{K_{m,k}+1}(K_{m,k}+\eta_{nk}),\hspace{0.2cm}\\&
\mathbb{E}[|\hat{h}_{nk}|^4]=\beta_{nk}^2\frac{(K_{m,k}^2 + 4K_{m,k} \eta_{nk}+2\eta_{nk}^2)}{(K_{m,k}+1)^2},\\&
\mathbb{E}[|\hat{h}_{nk}\hat{h}_{nj}|^2]=\beta_{nk}\beta_{nj}\frac{(K_{m,k}\eta_{nj} + K_{m,j}\eta_{nk}+ \eta_{nj}\eta_{nk})}{(K_{m,k} +1)(K_{m,j} +1)}, \hspace{0.2cm}\\&
\mathbb{E}[|\tilde{h}_{nj}|^2]=\frac{\beta_{nj}}{(K_{m,j}+1)(1+\tau_pp_\tau\beta_{nj})},\\&
\mathbb{E}[|\hat{h}_{nk}q_{ni}|^2]=\frac{\sigma^2_{m,ni}\beta_{nk}}{K_{m,k}+1}(K_{m,k}+\eta_{nk}).  
\end{align*}
By combining the terms and performing simple mathematical manipulations, we can obtain \eqref{eqrateanalmc1}. Please note that \eqref{eqrateanalsc1} and Lemma 2 are derived by following the strategy enumerated above.

% you can choose not to have a title for an appendix
% if you want by leaving the argument blank
%\section{}
%Appendix two text goes here.
%
%
%% use section* for acknowledgment
\section*{Acknowledgment}
The authors would like to express much appreciation to Prof. Andreas Molisch of University of Southern California. His comments and recommendations greatly contributed towards the successful publication of our paper.

% Can use something like this to put references on a page
% by themselves when using endfloat and the captionsoff option.
\ifCLASSOPTIONcaptionsoff
  \newpage
\fi

% trigger a \newpage just before the given reference
% number - used to balance the columns on the last page
% adjust value as needed - may need to be readjusted if
% the document is modified later
%\IEEEtriggeratref{8}
% The "triggered" command can be changed if desired:
%\IEEEtriggercmd{\enlargethispage{-5in}}

% references section

% can use a bibliography generated by BibTeX as a .bbl file
% BibTeX documentation can be easily obtained at:
% http://mirror.ctan.org/biblio/bibtex/contrib/doc/
% The IEEEtran BibTeX style support page is at:
% http://www.michaelshell.org/tex/ieeetran/bibtex/
%\bibliographystyle{IEEEtran}
% argument is your BibTeX string definitions and bibliography database(s)
%\bibliography{IEEEabrv,../bib/paper}
%
% <OR> manually copy in the resultant .bbl file
% set second argument of \begin to the number of references
% (used to reserve space for the reference number labels box)
%\begin{thebibliography}{1}
%
%\bibitem{IEEEhowto:kopka}
%H.~Kopka and P.~W. Daly, \emph{A Guide to \LaTeX}, 3rd~ed.\hskip 1em plus
%  0.5em minus 0.4em\relax Harlow, England: Addison-Wesley, 1999.
%
%\end{thebibliography}
\bibliographystyle{IEEEtr}
\end{document}